\documentclass[10pt,preprint]{aastex}
\usepackage{relsize}
\usepackage[colorlinks,urlcolor=blue,citecolor=blue,linkcolor=blue]{hyperref}
\usepackage{graphicx}
\shorttitle{UV luminosity function of high-z galaxies}
\shortauthors{Z.-Y. Cai et al.}

\begin{document}
\title{A physical model for the evolving UV luminosity function of high redshift galaxies and their contribution to the cosmic reionization}
\author
{Zhen-Yi~Cai\altaffilmark{1,2}, Andrea Lapi\altaffilmark{3,1}, Alessandro Bressan\altaffilmark{1}, Gianfranco De Zotti\altaffilmark{1,4}, Mattia Negrello\altaffilmark{4},  Luigi Danese\altaffilmark{1}}
\altaffiltext{1}{Astrophysics Sector, SISSA, Via Bonomea 265, I-34136 Trieste, Italy; zcai@sissa.it}
\altaffiltext{2}{Department of Astronomy and Institute of Theoretical Physics and Astrophysics, Xiamen University, Xiamen 361005, China}
\altaffiltext{3}{Dipartimento di Fisica, Universit\`a ``Tor Vergata'', Via della Ricerca Scientifica 1, I-00133 Roma, Italy}
\altaffiltext{4}{INAF - Osservatorio Astronomico di Padova, Vicolo dell'Osservatorio 5, I-35122 Padova, Italy}

\def\lsim{\,\lower2truept\hbox{${<\atop\hbox{\raise4truept\hbox{$\sim$}}}$}\,}
\def\gsim{\,\lower2truept\hbox{${> \atop\hbox{\raise4truept\hbox{$\sim$}}}$}\,}
\newcommand{\mum}{{\rm \mu m}}
\newcommand{\angstrom}{{\rm \mathring A}}
\newcommand{\cmsquare}{{\rm cm^2}}
\newcommand{\pcmcube}{{\rm cm^{-3}}}
\newcommand{\gram}{{\rm g}}
\newcommand{\Msun}{{M_\odot}}
\newcommand{\gpcmcube}{{\rm g\ cm^{-3}}}
\newcommand{\cmps}{{\rm cm\ s^{-1}}}
\newcommand{\ps}{{\rm s^{-1}}}
\newcommand{\pyr}{{\rm yr^{-1}}}
\newcommand{\yr}{{\rm yr}}
\newcommand{\Gyr}{{\rm Gyr}}
\newcommand{\ergspspHz}{{\rm ergs\ s^{-1}\ Hz^{-1}}}
\newcommand{\ergsps}{{\rm ergs\ s^{-1}}}

\begin{abstract}
We present a physical model for the evolution of the ultraviolet (UV) luminosity function of high redshift galaxies taking into account in a self-consistent way their chemical evolution and the associated evolution of dust extinction. {Dust extinction is found to increase fast with halo mass. A strong correlation between dust attenuation and halo/stellar mass for UV selected high-$z$ galaxies is thus predicted.} The model yields good fits of the UV and Lyman-$\alpha$ (Ly$\alpha$) line luminosity functions at all redshifts at which they have been measured. {The weak observed evolution of both luminosity functions between $z=2$ and $z=6$ is explained as the combined effect of the negative evolution of the halo mass function, of the increase with redshift of the star formation efficiency, due to the faster gas cooling, and of dust extinction, differential with halo mass. The slope of the faint end of the UV luminosity function is found to steepen with increasing redshift, implying that low luminosity galaxies increasingly dominate the contribution to the UV background at higher and higher redshifts. } The observed range of UV luminosities at high-$z$ implies a minimum halo mass capable of hosting active star formation $M_{\rm crit}\lesssim 10^{9.8}\,M_\odot$, consistent with the constraints from hydrodynamical simulations. {From fits of Ly$\alpha$ line luminosity functions plus data on the luminosity dependence of extinction and from the measured ratios of non-ionizing UV to Lyman-continuum flux density for samples of $z\simeq 3$ Lyman break galaxies and Ly$\alpha$ emitters, we derive a simple relationship between the escape fraction of ionizing photons and the star formation rate. It implies that the escape fraction is larger for low-mass galaxies, which are almost dust-free and have lower gas column densities. Galaxies already represented in the UV luminosity function ($M_{\rm UV}\lesssim -18$) can keep the universe fully ionized up to $z\simeq 6$. This is consistent with (uncertain) data pointing to a rapid drop of the ionization degree above $z\simeq 6$, such as indications of a decrease of the comoving emission rate of ionizing photons at $z\simeq 6$, of a decrease of sizes of quasar near zones, and of a possible decline of the Ly$\alpha$ transmission through the intergalactic medium at $z>6$}. On the other side, the electron scattering optical depth, $\tau_{\rm es}$, inferred from Cosmic Microwave Background (CMB) experiments favor {an ionization degree close to unity up to $z\simeq 9$--10. Consistency with CMB data can be achieved if $M_{\rm crit}\simeq 10^{8.5}\,M_\odot$, implying that the UV luminosity functions extend to $M_{\rm UV}\simeq -13$, although the corresponding $\tau_{\rm es}$ is still on the low side of CMB-based estimates.}

\end{abstract}



\section{Introduction}
\label{sec:intro}

One of the frontiers of present day astrophysical/cosmological research is the understanding of the transition from the ``dark ages'', when the hydrogen was almost fully neutral, to the epoch when stars and galaxies began to shine and the intergalactic hydrogen was almost fully re-ionized. {Observations with the Wide Field Camera 3 (WFC-3) on the Hubble Space Telescope \citep[HST,][]{Finkelstein2012,Bouwens2012b,Oesch2013a,Oesch2013b,Ellis2013,Robertson2013} have substantially improved the observational constraints on the abundance and properties of galaxies at cosmic ages of less than 1\,Gyr. Determinations of the ultraviolet (UV) luminosity function (LF) of galaxies at $z=7$--8 have been obtained by \citet{Bouwens2008,Bouwens2011b}, \citet{Smit2012}, \citet{Oesch2012}, \citet{Schenker2013}, \citet{McLure2010,McLure2013}, \citet{Yan2011,Yan2012}, \citet{Lorenzoni2011,Lorenzoni2013}, and \citet{Bradley2012}. Estimates over limited luminosity ranges were provided by \citet{Bouwens2008}, \citet{Oesch2013a}, and \citet{McLure2013} at $z=9$, and by \citet{Bouwens2011b} and \citet{Oesch2013a,Oesch2013b} at $z=10$.} Constraints on the UV luminosity density at redshifts up to 12 have been presented by \citet{Ellis2013}.

Since galaxies at $z\geqslant 6$ are the most likely sources of the UV photons capable of ionizing the intergalactic hydrogen, the study of the early evolution of the UV luminosity density is directly connected with the understanding of the cosmic reionization. Several studies \citep{Robertson2013,KuhlenFaucherGiguere2012,Alvarez2012,HaardtMadau2012} have adopted {\it parameterized} models for the evolving UV luminosity density. These models are anchored to the observed high-$z$ LFs and are used to investigate plausible reionization histories, consistent with other probes of the redshift-dependent ionization degree and primarily with the electron scattering optical depth measured by the Wilkinson Microwave Anisotropy Probe \citep[WMAP,][]{Hinshaw2013}.

{There are also several theoretical models for the evolution of the LFs of Ly$\alpha$ emitters (LAEs) and of Lyman break galaxies (LBGs), using different approaches. These include various semi-analytic galaxy formation models \citep{Tilvi2009,LoFaro2009,Kobayashi2010,Raicevic2011,Garel2012,Mitra2013,Lacey2011,GonzalezPerez2013}, smoothed particle hydrodynamics (SPH) simulations \citep{Dayal2009,Dayal2013,Nagamine2010,Salvaterra2011,Finlator2011b,Jaacks2012a,Jaacks2012b} as well as analytic models  \citep{Mao2007,Dayal2008,Samui2009,MunozLoeb2011,Munoz2012,SalimLee2012,Tacchella2013}. Each approach has known strengths and weaknesses. Given the complexity and the large number of variable parameters in many models, an analytic physical approach is particularly useful in understanding the role and the relative importance of the ingredients that come into play.  We adopt such approach, building on the work by \citet{Mao2007}. As mentioned above, spectacular advances in direct determinations of the UV LFs of galaxies at epochs close to the end of reionization have been recently achieved. These imply much stronger constraints on models, particularly on the analytic ones, that need to be revisited. Further constraints, generally not taken into account by all previous studies, have been provided by far-IR/(sub-)millimeter data, that probe other phases of the early galaxy evolution, concurring to delineate the complete picture.}

{A key novelty of the model is that it includes a self-consistent treatment of dust absorption and re-emission, anchored to the chemical evolution of the interstellar medium (ISM). This allows us to simultaneously account for the demography of both UV- and far-IR/(sub-)millimeter-selected high-$z$ star-forming galaxies. In addition, the model incorporates a variety of observational constraints that are only partially taken into account by most previous studies. Specifically, we take into account constraints on: the escape fraction of ionizing photons coming from both continuum UV and Ly$\alpha$ line LFs and from measurements of the ratio of non-ionizing to ionizing emission from galaxies; the luminosity/stellar mass--metallicity \citep{Maiolino2008,Mannucci2010} and the stellar mass--UV luminosity \citep{Stark2009,Stark2013} relations; the amplitude of the ionizing background at several redshifts up to $z=6$ \citep{DallAglio2009,WyitheBolton2011,Calverley2011,KuhlenFaucherGiguere2012,BeckerBolton2013,Nestor2013}. The successful tests of the model against a wide variety of data constitute a solid basis for extrapolations to luminosity and redshift ranges not yet directly probed by observations.}

The plan of the paper is the following. In Section\,\ref{sect:model}, we outline the model describing its basic ingredients. In Section\,\ref{sect:UV_LFs}, we exploit it to compute the cosmic-epoch dependent UV LF, allowing for the dust extinction related to the chemical evolution of the gas. In Section\,\ref{sect:Ionizing_luminosity}, we compute the production rate of ionizing photons and investigate their absorption rates by both dust and neutral hydrogen (HI), as constrained by measurements of the Ly$\alpha$ line LFs at various redshifts and of the ratios of non-ionizing UV to Lyman-continuum luminosities. We then derive the fraction of ionizing photons that can escape into the intergalactic medium (IGM) and use the results to obtain the evolution with redshift of the volume filling factor of the intergalactic ionized hydrogen. The main conclusions are summarized in Section\,\ref{sect:conclusions}.

Throughout this paper we adopt a flat $\Lambda \rm CDM$ cosmology with matter density $\Omega_{\rm m} = 0.32$, $\Omega_{\rm b} = 0.049$, $\Omega_{\Lambda} = 0.68$, Hubble constant $h=H_0/100\, \rm km\,s^{-1}\,Mpc^{-1} = 0.67$, spectrum of primordial perturbations with index $n = 0.96$, and normalization $\sigma_8 = 0.83$ \citep{PlanckCollaborationXVI2013}. All the quoted magnitudes are in the AB system \citep{OkeGunn1983}.



\section{Outline of the model}\label{sect:model}

{In this Section we present the basic features of the model used to compute the redshift-dependent UV and Lyman-$\alpha$  (Ly$\alpha$) LFs. The model is essentially the same used by \citet{Cai2013} in his study of the evolution of the IR LF. However, the \citet{Cai2013} study dealt with a later evolutionary phase, when star formation was dust-enshrouded. Here we are interested in the phase in which dust extinction was low but growing as the gas was being enriched in metals and the dust abundance was correspondingly increasing. The associated evolution with galactic age of dust extinction is discussed in Section~\ref{sect:gal_prop}. In Section~\ref{sect:test}, the model, integrated with this additional ingredient, is tested against data on high-$z$ galaxies not used to constrain the model parameters. }

\subsection{Basic ingredients}\label{sect:ingr}

Our approach is in line with the scenario worked out by \citet{Granato2004} and further elaborated by \citet{Lapi2006,Lapi2011}, \citet{Mao2007}, and \citet{Cai2013}. {It exploits the outcomes of many intensive $N$-body simulations and semi-analytic studies \citep[e.g.,][]{Zhao2003,Wang2011,LapiCavaliere2011} according to which pre-galactic halos undergo an early phase of fast collapse, including a few major merger events, during which the central regions reach rapidly a dynamical quasi-equilibrium; a subsequent slower accretion phase follows, which mainly affects the halo outskirts. We take the transition between the two phases as  the galaxy formation time.}

{The model envisages that, during the fast collapse phase, the baryons falling into the newly created potential wells are shock-heated to the virial temperature. This assumption may be at odds with analytic estimates and SPH simulations showing that the amount of shock-heated gas is relatively small for halo masses below $\sim 10^{12}\,M_\odot$ \citep{Keres2005,DekelBirnboim2006}, i.e., in the mass range of interest here (see below). However, the analytic estimates assume spherical collapse and the SPH simulations consider a smooth initial configuration. In both cases, the results do not necessarily apply to the halo build up during the fast collapse phase, dominated by a few major mergers. A crucial test of the star formation history implied by the model is provided by the data discussed in the following.}

{An alternative picture for the formation of galaxies is the ``cold stream driven'' scenario, according to which massive galaxies (halo masses above $\sim 10^{12}\,M_\odot$) formed from steady, narrow, cold gas streams, fed by dark matter filaments from the cosmic web, that penetrate the shock-heated media of massive dark matter haloes \citep{Dekel2009}. However, this mechanism may require an implausibly high star formation efficiency \citep{SilkMamon2012,Lapi2011} and a too high bias factor of galaxies at $z\simeq 2$ \citep{Dave2010,Xia2012}. Also observational evidence of cold flows has been elusive. The recent adaptive optics--assisted SINFONI observations of a $z=2.3285$ star-forming galaxy obtained at the Very Large Telescope \citep{Bouche2013} showed evidence of cold material at about one-third of the virial size of the halo. The gas appears to be co-rotating with the galaxy and may eventually inflow towards the center to feed star formation. However, this gas does not necessarily come from the intergalactic space; it may well be the result of cooling of the gas inside the halo, as implied by our scenario. Anyway, as we will see in the following, massive galaxies are of only marginal importance in the present context. }

\begin{figure*} 
\begin{center}
	\includegraphics[width=0.5\textwidth]{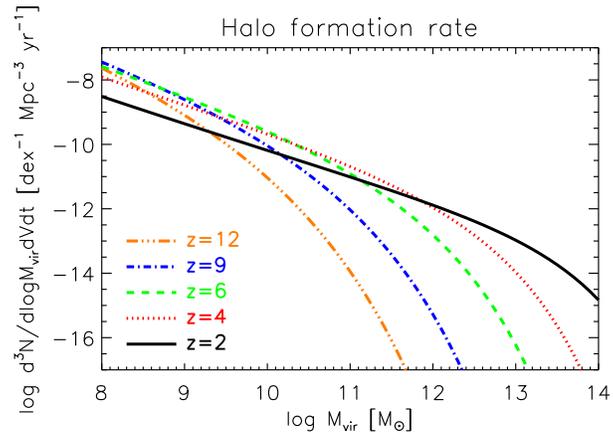}
	\caption{Evolution with redshift of the halo formation rate function. }\label{fig:halo_HFR}
\end{center}
\end{figure*}

In our framework, the galaxy LF is directly linked to the halo formation rate, $d^3 N(M_{\rm vir}, \tau)/ dM_{\rm vir}\, d\tau\,dV$, as a function of halo mass, $M_{\rm vir}$, and cosmic time, $\tau$ (see Figure~\ref{fig:halo_HFR}). The halo formation rate is approximated by the positive term of the derivative with respect to $\tau$ of the cosmological halo mass function \citep{Sasaki1994}. \citet{Lapi2013}, using an excursion set approach, showed that this is a good approximation. For the halo mass function, we used the analytic approximation by \citet{ShethTormen1999}.

In converting the halo formation rate into the UV LF, we also need to take into account the survival time, $t_{\rm destr}$, of the halos that are subject to merging into larger halos. {A short halo survival time would obviously decrease the number density of halos of given mass that are on at a given time. At the high redshifts of interest here, we expect that a large fraction of halos undergo major mergers, loosing their individuality; therefore the survival time could impact our estimate of the LF.}

The survival time is difficult to define unambiguously. We adopt the value obtained from the negative term {of the time derivative}, $(\partial_t N)_{-}$, of the \citet{PressSchechter1974} or of the \citet{ShethTormen1999} mass functions, i.e.,  $t_{\rm destr} \equiv N/ (\partial_t N)_{-}= t_{\rm Hubble}(z)$, independent of halo mass. {In our calculations of the UV LF (Equation~(\ref{LF_scatter1})) we cut the integration over time at $t_{\rm destr}$. As we will see in the following, in the mass and redshift ranges of interest here, the duration of the UV bright phase of galaxy evolution is generally substantially shorter than $t_{\rm Hubble}(z)$ and therefore also shorter than $t_{\rm destr}$. Therefore, the results are only weakly affected by the cut. Hence a more refined treatment of this tricky issue is not warranted.}

The history of star formation and of accretion into the central BH were computed by numerically solving the set of equations laid down in the Appendix of \citet{Cai2013}. These equations describe the evolution of the gas {along the following lines. Initially, a galactic halo of mass $M_{\rm vir}$ contains a cosmological gas mass fraction $f_{\rm b} = M_{\rm gas}/M_{\rm vir}=0.17$, distributed according to a \citet{Navarro1997} density profile, with a moderate clumping factor and primordial composition. The gas is heated to the virial temperature at the virialization redshift, $z_{\rm vir}$, then cools and  flows towards the central regions. The cooling is fast especially in the denser central regions, where it triggers a burst of star formation. We adopt a \citet{Chabrier2003} initial mass function (IMF). The gas cooling rate is computed adopting the cooling function given by \citet{SutherlandDopita1993}.  The evolution of the metal abundance of the gas is followed using the classical equations and stellar nucleosynthesis prescriptions, as reported, for instance, in \citet{Romano2002}.}

An appreciable amount of cooled gas loses its angular momentum via the drag by the stellar radiation field, settles down into a reservoir around the central supermassive black hole (BH), and eventually accretes onto it by viscous dissipation, powering the nuclear activity. The supernova (SN) explosions and the nuclear activity feed energy back to the gaseous baryons, and regulate the ongoing star formation and the BH growth.

{The feedback from the active nucleus has a key role in quenching the star formation in very massive halos but is only of marginal importance for the relatively low halo masses of interest here. This implies that the results presented in this paper are almost insensitive to the choice for the parameters governing the BH growth and feedback. The really important model parameters are only those that regulate the star formation rate (SFR), hence also the chemical evolution. These are the clumping factor $C$, that affects the cooling rate, the ratio between the  gas inflow timescale, $t_{\rm cond}$, and the star formation timescale, $t_\star$, $s=t_{\rm cond}/t_\star$, and the SN feedback efficiency, $\epsilon_{\rm SN}$.  }

{For the first two of the latter parameters, the clumping factor $C$ and ratio $s$, as well as for all the parameters controlling the growth of and the feedback from the active nucleus that, albeit almost irrelevant, are nevertheless included in our calculations, we adopt the values determined by \citet{Lapi2006} and also used by \citet{Lapi2011} and \citet{Cai2013}, i.e., $C=7$ and $s=5$. A discussion on the plausible ranges for all the model parameters can be found in \citet{Cai2013}.}

{On the other hand, we have made a different choice for $\epsilon_{\rm SN}$. This is motivated by the fact that the earlier papers dealt with relatively high halo masses ($M_{\rm vir}\gtrsim 10^{11.2}\ M_\odot$) while, at the high redshifts of interest here, the masses of interest are substantially lower.} As pointed out by \citet{Shankar2006}, the data indicate that the star formation efficiency in low-mass halos must be lower than implied by the SN feedback efficiency gauged for higher halo masses. In other words, as extensively discussed in the literature, external (UV background) or internal (SN explosions, radiation from massive low-metallicity stars and, possibly, stellar winds) heating processes can reduce the SFR in low-mass halos {\citep[e.g.,][]{PawlikSchaye2009,Hambrick2011,Finlator2011a,KrumholzDekel2012,Pawlik2013,WyitheLoeb2013,SobacchiMesinger2013}} and completely suppress it below a critical halo mass, $M_{\rm crit}$.

We take this into account by increasing the SN feedback efficiency with decreasing halo mass and introducing a low-mass cutoff, i.e., considering only galaxies with $M_{\rm vir}\geqslant M_{\rm crit}$. We set
\begin{equation}\label{eq:epsilon}
\epsilon_{\rm SN} = 0.05 \{3 + {\rm erf}[-\log(M_{\rm vir}/10^{11}\,M_\odot)/0.5]\} / 2,
\end{equation}
that converges, for $M_{\rm vir}> 10^{11}\,M_\odot$, to the constant value, $\epsilon_{\rm SN} = 0.05$, used by \citet{Cai2013}.

The critical halo mass can be estimated equating the gas thermal speed in virialized halos to the escape velocity to find $M_{\rm crit}\simeq 2\times 10^8(T/2\times 10^4\,\hbox{K})^{3/2}\,M_\odot$ \citep{Hambrick2011}, where $2\times 10^4\,\hbox{K}$ corresponds to the peak of the hydrogen cooling curve. If the balance between heating and cooling processes results in higher gas temperatures, $M_{\rm crit}$ can increase even by a large factor. An accurate modeling of the physical processes involved is difficult and numerical simulations yield estimates of $M_{\rm crit}/M_\odot$ in the range from $4\times 10^8$ to $\gtrsim 10^9$ \citep{Hambrick2011,KrumholzDekel2012,Pawlik2013,SobacchiMesinger2013}, with a trend toward lower values at $z >6$ {when the IGM was cooler \citep[e.g.,][]{Dijkstra2004,Kravtsov2004,Okamoto2008} and the intensity of the UV background is expected to decrease.}

If, on one side, heating properties can decrease the SFR in small galactic halos, the production of UV photons per unit stellar mass is expected to substantially increase with redshift, at least at $z\gtrsim 3$.  \citet{Padoan1997} argued that the characteristic stellar mass, corresponding to the peak in the IMF per unit logarithmic mass interval, scales as the square of the temperature, $T_{\rm mc}$, of the giant molecular clouds within which stars form. But the typical local value of $T_{\rm mc}$ ($\sim 10\,$K) is lower than the cosmic microwave background temperature at $z\gtrsim 3$, implying a rapid increase with $z$ of the characteristic stellar mass, hence of the fraction of massive stars producing UV photons. On the other hand, a higher fraction of massive stars also implies a more efficient gas heating, hence a stronger quenching of the SFR. This complex set of phenomena tend to counterbalance each other. Therefore, in the following we adopt, above $M_{\rm crit}$, the SFR given by the model and the UV (ionizing and non-ionizing) photon production rate appropriate to the  \citet{Chabrier2003} IMF, allowing for the possibility of a correction factor to be determined by comparison with the data.

\subsection{Evolution of galaxy properties}\label{sect:gal_prop}

{The model outlined above allows us to compute the evolution with galactic age of the SFR, of the mass in stars, and of the metal abundance of the gas as a function of the halo mass, $M_{\rm vir}$, and of the halo virialization redshift, $z_{\rm vir}$. Illustrative examples  for four values of the halo mass at fixed $z_{\rm vir}=6$ (left panels) and for four values of $z_{\rm vir}$ and two values of the halo mass (right panels) are shown in Figure~\ref{fig:sph_evol_oqs1}. A few points are worth noting: i) the active star formation phase in the most massive halos ($M_{\rm vir}\gtrsim 10^{12}\,M_\odot$) is abruptly terminated by the AGN feedback, while in less massive halos, where the feedback is dominated by stellar processes, the SFR declines more gently; ii) at fixed halo mass, the SFR is initially higher at higher redshifts (mainly because the higher densities imply faster cooling of the gas) and declines earlier; iii) the higher star formation efficiency at higher $z$ implies that the stellar to halo mass ratio and the metallicity are higher for galaxies that formed at higher redshifts. The different behavior of these quantities for different halo masses and different redshifts determines the relative contributions of the various masses to the LFs and their evolution with cosmic time, as discussed in the following.}

{For comparisons with the data we further need to take into account the dust extinction. The latter is proportional to the dust column density which is proportional to the gas column density. In turn, the dust column density is proportional to the gas column density, $N_{\rm gas}$, times the metallicity, $Z_{\rm g}$, to some power reflecting the fraction of metals locked into dust grains. We thus expect that at our reference UV wavelength, 1350 $\angstrom$, the dust extinction $A_{1350}\propto \dot{M}_{\mathrm{\star}}^{\alpha} Z_{\rm g}^\beta$, where $\dot M_{\mathrm{\star}}$ is the star formation rate. \citet{Mao2007} found that a relationship of this kind ($A_{1350}\approx 0.35\,(\dot{M}_{\star}/M_{\odot}\,\mathrm{yr}^{-1})^{0.45}\,(Z_{\rm g}/Z_{\odot})^{0.8}$) does indeed provide a good fit of the luminosity-reddening relation found by \citet{Shapley2001}. We have adopted a somewhat different relation
\begin{equation}\label{eq|extgigi}
A_{1350}\approx 0.75\,\left(\frac {\dot{M}_{\star}}{M_{\odot}\,\mathrm{yr}^{-1}}\right)^{0.25}\, \left(\frac{Z_{\rm g}}{Z_{\odot}}\right)^{0.3},
\end{equation}
that we have found to provide a better fit of the UV LFs of LBGs \citep[][see below]{ReddySteidel2009,Bouwens2007,McLure2013} still being fully consistent with the \citet{Shapley2001} data (cf. Figure~\ref{fig:M_EBV}) as well as with observational estimates of the escape fractions ($f^{\rm esc}_\lambda=\exp(-A_\lambda/1.08)$, see below) of UV and Ly$\alpha$ photons (Figure~\ref{fig:fesc_UVLya}).  }

{As illustrated by the bottom-left panel of Figure~\ref{fig:sph_evol_oqs1}, the UV extinction is always low for the least massive galaxies ($M_{\rm vir}\lesssim 10^{11}\,M_\odot$), but increases rapidly with increasing $M_{\rm vir}$. This implies a strong correlation between dust attenuation and halo/stellar mass, thus providing a physical explanation for the correlation reported by \citet[][cf. their Figure~3]{Heinis2014}. The bottom-right panel of the figure shows that, at fixed halo/stellar mass, the model implies a mild increase of the attenuation with increasing $z$, paralleling the small increase in the gas metallicity due to the higher star formation efficiency.}

{The observed absolute magnitude in the AB system of a galaxy at our reference UV wavelength, $\lambda = 1350\,${\AA},  is related to its SFR and to dust extinction $A_\lambda$ as
\begin{equation}\label{eq:magn1350}
M_{1350}^{\rm obs}=51.59 - 2.5\, \log{L_{1350}^{\rm int}\over \mathrm{erg\,s}^{-1}\,\mathrm{Hz}^{-1}} + A_{1350},
\end{equation}
with
\begin{equation}\label{eq:UV2SFR_Cha}
{L_{1350}^{\rm int}} = k_{\rm UV} {\dot M_\star},
\end{equation}
where $L_{1350}^{\rm int}$ is the intrinsic monochromatic luminosity at the frequency corresponding to 1350\,{\AA}.  On account of the uncertainties on the IMF at high $z$, mentioned above, we treat the normalization factor $k_{\rm UV}\, (\hbox{erg}\, \hbox{s}^{-1}\, \hbox{Hz}^{-1}\, M_\odot^{-1}\, \hbox{yr})$ as an adjustable parameter. Using the evolutionary stellar models of \citet{Fagotto1994a,Fagotto1994b}, we find that $k_{\rm UV}$ varies with galactic age and gas metallicity as illustrated by Figure~\ref{fig:t_sfr_z}. The main contributions to the UV LF come from galactic ages $\gtrsim 10^7\,$yr, when $\log(k_{\rm UV})\gtrsim 27.8$ and is somewhat higher for lower metallicity. We adopt $\log(k_{\rm UV}) = 28$ as our reference value.
We remind that the relationship between the absolute AB magnitude at the effective frequency $\nu$, $M_{\nu}$, and the luminosity is $\log[(\nu\,L_\nu)/L_\odot]=2.3995-\log(\lambda/1350\,\angstrom)-0.4\,M_{\nu}$ where $L_\odot=3.845\times 10^{33}\,\hbox{erg}\,\hbox{s}^{-1}$.}

{We adopt the standard ``dust screen'' model for dust extinction, so that the observed luminosity is related to the intrinsic one by
\begin{equation}
	L^{\rm obs}(\lambda) = L^{\rm int}(\lambda) 10^{-0.4A_\lambda},
\end{equation}
with the \citet{Calzetti2000} reddening curve, holding for $0.12\,\mum\, \leqslant \lambda \leqslant 0.63\,\mum$,
\begin{equation}\label{eq:Calzetti}
	A_\lambda/E_s(B-V) \simeq 2.659 (-2.156 + 1.509/\lambda - 0.198/\lambda^2 + 0.011/\lambda^3) + R_V
\end{equation}
where $E_s(B-V)$ is the color excess of the galaxy stellar continuum and $R_V = 4.05$. This gives $A_{1350}/E_s(B-V) \simeq 11.0$ and $A_{1216} \simeq 1.08 A_{1350}$. \citet{Oesch2013c} found that the dust reddening at $z \simeq 4$ is better described by a Small Magellanic Cloud (SMC) extinction curve \citep{Pei1992} rather than by the Calzetti curve. However, for our purposes the only relevant quantity is the $A_{1216}/A_{1350}$ ratio. Using the \citet{Pei1992} fitting function for the SMC curve we find $(A_{1216}/A_{1350})_{\rm SMC}=1.14$, very close to the Calzetti ratio (1.08). Thus our results would not appreciably change adopting the SMC law. }

\subsection{Testing the model against high-\texorpdfstring{$z$}{} data}\label{sect:test}

{Some average properties, weighted by the halo formation rate, of galaxies at $z_{\rm obs}=2$, 6, and 12 are shown, as a function of the observed luminosity, in Figures~\ref{fig:properties_LBGs_LAEs} and \ref{fig:Zg_Mstar}. The weighted averages have been computed as follows. We sample the formation rate of halos with mass $M_{\rm vir}$ virializing at $t_{\rm vir}$, $\dot n(M_{\rm vir}, t_{\rm vir}) \equiv d^3N(M_{\rm vir}, t_{\rm vir})/d\log M_{\rm vir}\,dt\,dV$, in steps of $\Delta \log M_{\rm vir} = 0.08$ and cosmic time interval $\Delta t_{\rm vir} = 4\,$Myr, and select those having some property $X$, like the UV, $L_{1350}$ luminosity; the Ly$\alpha$, $L_{1216}$, luminosity; or the stellar mass (actually we set $X=\log(\hbox{quantity})$) in a given range, $X_i \in (X - \Delta X/2, X + \Delta X/2]$, at the selected $z_{\rm obs}$. The average value, $\mu_Y(X, \Delta X)$, of some other property $Y$ (e.g., age of stellar populations, $t_{\rm age}$, gas metallicity, $Z_{\rm g}$, SFR, etc.) is then computed as
\begin{equation}
	\mu_Y(X, \Delta X) = \frac{\sum_i Y_i \,\dot n_i \, \theta_{\rm H}(X - \Delta X / 2 < X_i \leqslant X + \Delta X / 2)}{ \sum_i \dot n_i \, \theta_{\rm H}(X - \Delta X / 2 < X_i \leqslant X + \Delta X / 2)},
\end{equation}
where $\theta_{\rm H}$ ($\theta_{\rm H} (x)= 1\ \mathrm{if}\ x\ \mathrm{is\ true}$, $\theta_{\rm H} (x)= 0\ \mathrm{otherwise}$) is the Heaviside function. The comoving number density of galaxies satisfying the selection criterion is
\begin{equation}
	n(X, \Delta X) = \frac{1}{\Delta X} \sum_i \dot n_i  \cdot \Delta \log M_{\rm vir} \cdot \Delta t_{\rm vir} \cdot \theta_{\rm H}(X - \Delta X / 2 < X_i \leqslant X + \Delta X / 2).
\end{equation}
}%
{The data we compare with are specified in the panels and in the captions of the figures. Note that these data were \textit{not} used to constrain the model parameters. As illustrated by Figure~\ref{fig:properties_LBGs_LAEs}, the model compares favorably with observational estimates of various galaxy properties at different redshifts. Stellar population ages are, on average, substantially younger for the brighter galaxies, which are associated to the most massive halos. This is fully consistent with the well established ``downsizing'' scenario since it happens because massive galaxies have high SFRs that may reach $\hbox{SFR}\sim 100\,M_\odot\,\hbox{yr}^{-1}$ and therefore develop their stellar populations faster. The chemical enrichment and the associated dust obscuration grow also faster in these galaxies that become soon UV-faint. This causes a downward trend of $t_{\rm age}$ with increasing $L^{\rm obs}_{1350}$, consistent with  the data (upper left panel of Figure~\ref{fig:properties_LBGs_LAEs}).}

{To compare the predictions of our model with observations on the metallicity--luminosity and metallicity--stellar mass relations at different redshifts, we have converted the values of $[12 + \log({\rm O/H})]$ given by \citet{Maiolino2008} and \citet{Liu2008} to $Z_{\rm g}$ in solar units adopting a solar oxygen abundance $[12 + \log({\rm O/H})]_\odot = 8.69$ \citep{Asplund2009}. In Figure~\ref{fig:Zg_Mstar} the stellar masses estimated by \citet{Maiolino2008} assuming a Salpeter IMF have been divided by a factor of 1.4 to convert them to the Chabrier IMF used in the model. Also we have plotted the metallicity estimates by \citet{Erb2006} for a sample of 87 LBGs at $z \sim 2.2$ as revised by \citet{Maiolino2008} using an improved calibration. \citet{Cullen2013} found an offset in the relation for $z\sim 2$ galaxies compared to \citet{Erb2006}. They argue that the difference originates from the use of different metallicity estimators with locally calibrated metallicity relations that may not be appropriate for the different physical conditions of star-forming regions at high redshifts. }

{The dependence of gas-phase metallicity on stellar mass, with lower $M_\star$ galaxies having lower metallicities, is accounted for by the model, at least at $z_{\rm obs}=2.2$ (Figure~\ref{fig:Zg_Mstar}). This trend is also present in the data on $z_{\rm obs}=3.4$ galaxies by \citet{Maiolino2008} which indicate a decrease by 0.6 dex of the amplitude of the $Z_{\rm g}$--$M_\star$ relation compared to that observed at $z_{\rm obs}=2.2$. Such strong evolution is at odds with model predictions, according to which the evolution should be quite weak, as shown in the bottom-right panel of Figure~\ref{fig:Zg_Mstar}. However, as argued by \citet{Cullen2013}, the strong evolution may be an artifact due to the use of locally calibrated metallicity indicators which do not account for evolution in the physical conditions of star-forming regions.}

\section{The redshift-dependent UV luminosity function}\label{sect:UV_LFs}

{In the previous Section we have introduced all the ingredients needed to compute the population properties of high-$z$ galaxies. We now proceed with the calculation of the cosmic epoch-dependent UV LF and discuss the role of dust extinction in shaping it (Section~\ref{sect:ext}), the constraints on its faint end (Section~\ref{sect:faint}) that, as we shall see, is important in the context of understanding the cosmic reionization, its cosmological evolution (Section~\ref{sect:evol}), and the transition from the dust-free to the dust-enshrouded star formation phases (Section~\ref{sect:UV_IR}). }

{The comoving differential UV LF $\Phi(\log L^{\rm obs}_{1350}, z)$, i.e., the number density of galaxies per unit $\log L^{\rm obs}_{1350}$  interval at redshift $z$, can be computed coupling the halo formation rate with the relationship between halo mass and SFR as a function of cosmic time, $\tau$, and with the relationship between SFR and UV luminosity (Equation~(\ref{eq:magn1350})) to obtain}
\begin{equation}\label{LF_scatter1}
\Phi(\log L^{\rm obs}_{1350}, z)\!\! =\!\!\!  \int^{M^{\rm max}_{\rm vir}}_{M^{\rm min}_{\rm vir}}\!\!\!\!\!  dM_{\rm vir} \int^{z^{\rm max}_{\rm vir}}_z\!\!\!\!\!\! dz_{\rm vir} \Big|\frac{d \tau_{\rm vir}}{dz_{\rm vir}}\Big| {d^3 N\over d M_{\rm vir}\, d \tau_{\rm vir}\, dV } \,  \theta_{\rm H}\bigl[t(z) - t(z_{\rm vir}) < t_{\rm destr}(z_{\rm vir})\bigr],
\end{equation}
%
%
%
{where $L^{\rm obs}_{1350}$ is the observed luminosity, attenuated by dust, and $\theta_{\rm H}(x)$ is the Heaviside function.
We set  $z^{\rm max}_{\rm vir}=16$, $M^{\rm max}_{\rm vir}= 10^{13.3}\,M_\odot$, and $M^{\rm min}_{\rm vir} = M_{\rm crit}$. 
}
%
%

%
%


Figure~\ref{fig:LF_LBGs_1350} shows that the model is nicely consistent with observational estimates of the UV LF  over the full redshift range from $z \sim 3$ to $z \sim 10$ for $k_{\rm UV}=1.0\times 10^{28}\,\hbox{erg}\,\hbox{s}^{-1}\,\hbox{Hz}^{-1}\,M_\odot^{-1}\,\hbox{yr}$  (Equation~(\ref{eq:UV2SFR_Cha}), see also Figure~\ref{fig:t_sfr_z}). Observational determinations of UV LFs were made at various rest-frame wavelengths. However, since the UV continuum slope, $\beta$ ($f_\lambda \propto \lambda^\beta$), of high-$z$ galaxies is close to $\beta=-2$ \citep{Bouwens2012b,Castellano2012}, the color correction $M_{\rm AB, \lambda 1}-M_{\rm AB,\lambda 2}=-2.5(\beta+2)\log(\lambda_1/\lambda_2)$ is small and has been neglected.


\subsection{The role of dust extinction}\label{sect:ext}

{Dust extinction is differential in luminosity as is clearly seen comparing the solid black lines (no extinction) in Figure~\ref{fig:LF_LBGs_1350} with the dot-dashed blue lines (extinction included). This follows from the faster metal enrichment in the more massive objects (see Figure~\ref{fig:sph_evol_oqs1}), where the SN and radiative feedbacks are less effective in depressing the SFR. }

{Figure~\protect\ref{fig:sph_evol_oqs1} also shows that the least massive galaxies ($M_{\rm vir} < 10^{11}\,M_\odot$) have low extinction throughout their lifetime. In conjunction with the fast decline of the number density of more massive halos, this results in an increasing dominance of the contribution of low mass galaxies to the UV LF at higher and higher $z$.  Figure~\ref{fig:LF_LBGs_1350} indeed shows that the observed portion of the UV LF at $z\geqslant 7$ is accounted for by galaxies with $M_{\rm vir} < 10^{11.2}\,M_\odot$, and the effect of dust extinction is negligible over most of the observed luminosity range. This offers the interesting possibility of reconstructing the halo formation rate from the UV LF without the complication of uncertain extinction corrections. }

\subsection{The faint end of the UV luminosity function}\label{sect:faint}

{The LF has, at the faint end, a break corresponding to $M_{\rm crit}$, the mass below which the star formation is suppressed by external and internal heating processes (see Section\,\ref{sect:model}). In Figure~\ref{fig:LF_LBGs_1350} $M_{\rm crit}=10^{8.5}\,M_\odot$ but in the panels comparing the model with the data at various redshifts the thin dotted and solid blue lines show the effect of increasing $M_{\rm crit}$ to $10^{9.8}\,M_\odot$ and $10^{11.2}\,M_\odot$, respectively. This shows that the data only constrain $M_{\rm crit}$ to be $\lesssim 10^{9.8}\,M_\odot$, consistent with estimates from simulations, summarized in Section~\ref{sect:ingr}. }

{Interestingly, \citet{MunozLoeb2011}, using a substantially different model, find that the observed UV LFs at $z=6$, 7, and 8 are best fit with a minimum halo mass per galaxy $\log(M_{\rm min}/M_\odot)= 9.4 (+0.3,-0.9)$ in good agreement with our estimate. These authors simply assume negligible dust extinction at these redshifts; our model provides a quantitative physical account of the validity of this assumption. \citet{Raicevic2011}, using the Durham GALFORM semi-analytical galaxy formation model, also find that the minimum halo mass that contributes substantially to the production of UV photons at these redshifts is $\log(M_{\rm min}/{h}^{-1}\,M_\odot) \simeq 9$. A lower value, $\log(M_{\rm min}/\,M_\odot) \simeq 8.2$, was obtained by \citet{Jaacks2012a} from their cosmological hydrodynamical simulations. \citet{Trenti2010} derive from abundance matching that the galaxies currently detected by {\it HST} live in dark matter halos with $M_{\rm H} \gtrsim 10^{10}\,M_\odot$, and they predict a weak decrease of the star formation efficiency with decreasing halo mass down to a minimum halo mass for star formation $M_{\rm H} \sim 10^8\,M_\odot$, under the assumption that the luminosity function remains a steep power law at the faint end.}

{The higher feedback efficiency for less massive halos (Equation~(\ref{eq:epsilon})) makes the slope of the faint end of the UV LF (above the break) flatter than that of the low mass end of the halo formation rate function  (illustrated by Figure~\ref{fig:halo_HFR}). However the former slope reflects to some extent the steepening with increasing redshift of the latter. Just above the low-luminosity break corresponding to $M_{\rm crit}$, the slope $-d\log \Phi(L^{\rm obs}_{1350},z)/d\log L^{\rm obs}_{1350}$ (where $\Phi$ is per unit $d\log L^{\rm obs}_{1350}$) steadily increases from $\simeq 0.5$ at $z=2$, to $\simeq 0.7$ at $z=6$, and to $\simeq 1$ at $z=10$ (see the bottom-right panel of Figure~\ref{fig:LF_LBGs_1350}). The steepening of the faint end of the UV LF with increasing $z$ implies an increasing contribution of low luminosity galaxies to the UV background. The dust extinction, differential with halo mass, further steepens the bright end of the observed UV LF.}

\subsection{Evolution of the UV luminosity function}\label{sect:evol}

{As shown by the bottom-right panel of Figure~\ref{fig:LF_LBGs_1350}, the model implies that the evolution of the UV LF from $z=2$ to $z=6$ is weak. In particular the fast decrease with increasing redshift of the high mass halo formation rate (Figure~\ref{fig:halo_HFR}) is not mirrored by the bright end of the LF. This is partly due, again, to the fast metal enrichment of massive galaxies that translates into a rapid increase of dust extinction and, consequently, in a short duration of their UV bright phase (Figure~\ref{fig:sph_evol_oqs1}). Thus the contribution to the UV LF of galaxies in halos more massive than $\sim 10^{11.2}\, M_\odot$ (thin solid blue line in Figure~\ref{fig:LF_LBGs_1350}) decreases rapidly with increasing redshift and galaxies with $M_{\rm vir} \gg 10^{11.2}\, M_\odot$ contribute very little. The UV LF at high $z$ is therefore weakly sensitive to the rapid variation of the formation rate of the latter galaxies. The milder decrease of the formation rates of less massive galaxies is largely compensated by the increase of the star formation efficiency due to the faster gas cooling (see Figure~\ref{fig:sph_evol_oqs1}). At $z>6$, however, the decrease of the formation rate even of intermediate mass galaxies prevails and determines a clear negative evolution of the UV LF.}



\subsection{From dust-free to dust-enshrouded star formation}\label{sect:UV_IR}

{As illustrated by the bottom-left panel of Figure~\ref{fig:sph_evol_oqs1}, the star formation in massive galaxies is dust enshrouded over most of its duration. For example, the figure shows that the extinction at 1350\,{\AA} of a galaxy with $M_{\rm vir}=10^{12}\,M_\odot$ steeply grows already at galactic ages of a few times $10^7\,$yr, reaching two magnitudes at an age of $4\times 10^{7}\,$yr. The total duration of the star formation phase for these galaxies is $\simeq 7\times 10^8[(1+z)/3.5]^{-1.5}\,$yr \citep{Cai2013}, which implies that they show up primarily at far-IR/(sub-)millimeter wavelengths. Thus the present framework entails a continuity between the high-$z$ galaxy SFR function derived from the UV LF (dominated by low-mass galaxies) and that derived from the infrared (IR; 8--$1000\,\mu$m) LF (dominated by massive galaxies), investigated by \citet{Lapi2011} and \citet{Cai2013}. This continuity is borne out by observational data at $z = 2$ and 3, as shown by Figure~\ref{fig:SFRF_z}. Once proper allowance for the effect of dust attenuation is made, the model accurately matches the IR and the UV LFs and, as expected, UV and IR data cover complementary SFR ranges.}

{The SFR histories inferred from UV and far-IR data are compared in Figure~\ref{fig:SFRD_8d5_18d0} \citep[see also][]{Fardal2007}. This figure shows that the ratio of dust-obscured to unobscured SFR increases with increasing redshift until it reaches a broad maximum at $z\sim 2$--3 and decreases afterwards. This entails a prediction for the evolution of the IR luminosity density, $\rho_{\rm IR}$, beyond $z\simeq 3.5$, where it cannot yet be determined observationally. According to the present model, at these redshifts, $\rho_{\rm IR}$ decreases with increasing $z$ faster than the UV luminosity density, $\rho_{\rm UV}$. In other words, the extinction correction needed to determine the SFR density from the observed $\rho_{\rm UV}$ is increasingly small at higher and higher redshifts. This is because the massive halos, that can reach high values of dust extinction, become increasingly rare at high $z$.}

\section{Ionizing photons}\label{sect:Ionizing_luminosity}

{We now move from the non-ionizing to the ionizing UV. Two issues need to be addressed: the production rate of ionizing photons as a function of halo mass and galactic age and the fraction of them that manages to escape into the IGM, contributing to its ionization rate. Key information on both issues is provided by the Ly$\alpha$ line LF, observationally determined up to $z\simeq 8$ with some constraints also at $z\simeq 9$. As discussed in Section~\ref{sect:LAE}, the observed Ly$\alpha$ luminosity of a galaxy is directly proportional to the production rate of ionizing photons, to their absorption fraction by HI in the ISM, and to the fraction of them that escapes absorption by dust. It is thus a sensitive probe of the rate at which such photons can escape into the IGM, ionizing it. Equipped with the information obtained from the redshift-dependent Ly$\alpha$ line LFs and taking into account recent measurements of the ratio of non-ionizing UV to Lyman-continuum emission, in Section~\ref{sect:esc} we evaluate the injection rate of ionizing photons into the IGM and discuss the cosmic reionization. This is done for several choices, taken from the literature, of the IGM clumping factor, an ingredient not provided by the model. The results are discussed vis-a-vis with a variety of observational constraints, including those from the electron scattering optical depth measured by CMB experiments.}

\subsection {Ly\texorpdfstring{$\alpha$}{} emitters}\label{sect:LAE}

The ionizing photons ($\lambda \leqslant 912\,\angstrom$) can be absorbed by both dust and HI. About $2/3$ of those absorbed by HI are converted into Ly$\alpha$ photons \citep{Osterbrock1989,Santos2004}, so that the LFs of LAEs provide information on their production rate. The Ly$\alpha$ line luminosity before extinction is then
\begin{equation}\label{eq:Llyaint}
L^{\mathrm{int}}_{{\mathrm{Ly}}\alpha}= {2\over 3}\,\dot{N}^{\mathrm{int}}_{912 }\,f^{\rm dust}_{912}\,(1-f^{\rm HI}_{912})h_{\rm P}\nu_{{\mathrm{Ly}}\alpha}
\end{equation}
where $\dot{N}^{\mathrm{int}}_{912 }$ is the production rate of ionizing photons while $f^{\rm dust}_{912}=\exp(-\tau_{\rm dust,ion})$ and $f^{\rm HI}_{912}=\exp(-\tau_{\rm HI})$ are the fractions of them that escape absorption by dust and by HI, respectively, and $h_{\rm P}$ is the Planck constant. {We model $\tau_{\rm HI}$ as
\begin{equation}\label{eq:tauHI}
	\tau_{\rm HI} = \tau^0_{\rm HI} \left( \frac{\dot M_\star}{M_\odot\,\rm yr^{-1}} \right)^{\alpha_{\rm HI}} + \beta_{\rm HI},
\end{equation}
where the first term on the right-hand side refers to the contribution from high density star-forming regions while the second term refers to the contribution of diffuse HI. Since the \citet{Calzetti2000} law does not provide the dust absorption optical depth of ionizing photons, we set
\begin{equation}\label{eq:tau_dust_ion}
\tau_{\rm dust,ion}=A_{912}/1.08=\gamma\,A_{1350}/1.08
\end{equation}
where $A_{1350}$ is given by Equation~(\ref{eq|extgigi}) and $\gamma$ is a parameter of the model.}

The evolution with galactic age of $\dot{N}^{\mathrm{int}}_{912 }$, for a \citet{Chabrier2003} IMF, a constant SFR of $1\,M_\odot\,\hbox{yr}^{-1}$, and three values of the gas metallicity is illustrated by Figure~\ref{fig:kion_Rint}. We adopt an effective ratio $k_{\rm ion}\equiv \dot N^{\rm int}_{912}/\dot M_\star = 4.0 \times 10^{53}\, \hbox{photons}\, \hbox{s}^{-1}\, M^{-1}_\odot\ \yr$, appropriate for the typical galactic ages and metallicities of our sources.  The figure also shows the evolution of the intrinsic ratio of Lyman-continuum ($\lambda \leqslant 912\,\angstrom$; $L^{\rm int}_{912} = \dot N^{\rm int}_{912} h_{\rm P} \rm\ erg\ s^{-1}\ Hz^{-1}$) to UV luminosity at 1350\,\AA, $R_{\rm int}\equiv L^{\rm int}_{912}/L^{\rm int}_{1350}$. For our choice of $k_{\rm ion}$ and $k_{\rm UV}$ (Equation~(\ref{eq:UV2SFR_Cha})) we have $R_{\rm int} \simeq 0.265$.  

The interstellar dust attenuates $L^{\mathrm{int}}_{{\mathrm{Ly}}\alpha}$ by a factor $f^{\rm dust}_{{\rm Ly}\alpha}= \exp(-\tau^{\mathrm{dust}}_{{\mathrm{Ly}}\alpha})$, where $\tau^{\mathrm{dust}}_{{\mathrm{Ly}}\alpha}$ is the dust optical depth at the Ly$\alpha$ wavelength (1216\,\AA). The physical processes governing the escape of Ly$\alpha$ photons from galaxies are complex. Dust content, neutral gas kinematics, and the geometry of the neutral gas seem to play the most important roles. For objects with low dust extinction, such as those relevant here, the detailed calculations of the Ly$\alpha$ radiation transfer by \citet[][see their Figures~18 and 19]{Duval2014} show that the Ly$\alpha$ is more attenuated than the nearby UV continuum by a factor $\simeq 2$, consistent with the observational result \citep[e.g.,][]{Gronwall2007} that the SFRs derived from the Ly$\alpha$ luminosity are $\simeq 3$ times lower than those inferred from the rest-frame UV continuum. With reference to the latter result, it must be noted that part of the attenuation of the Ly$\alpha$ luminosity must be ascribed to the IGM (see below) and that the discrepancy between Ly$\alpha$  and UV continuum SFRs may be due, at least in part, to uncertainties in their estimators.
A good fit of the Ly$\alpha$ line LFs is obtained setting $f^{\rm dust}_{{\rm Ly}\alpha}\simeq f^{\rm dust}_{1216}/1.6$, where $f^{\rm dust}_{1216} = \exp(-A_{1216}/1.08)$, consistent with the above results.  

The fraction $f^{\mathrm{IGM}}_{{\mathrm{Ly}}\alpha}=\exp(-\tau^{\mathrm{IGM}}_{{\mathrm{Ly}}\alpha})$ of Ly$\alpha$ photons that survive the passage through the IGM was computed following \citet{Madau1995}, taking into account only the absorption of the blue wing of the line
\begin{equation}\label{eq:Madau}
f^{\rm IGM}_{\rm Ly\alpha} = 0.5 \{1 + \exp[-0.0036(1+z)^{3.46}]\}.
\end{equation}
The strong attenuation by dust within the galaxy and by HI in the IGM implies that estimates of the SFR of high-$z$ LBGs and LAEs from the observed Ly$\alpha$ luminosity require careful corrections and are correspondingly endowed with large uncertainties. Vice versa, the statistics of LAEs and LBGs at high redshifts are sensitive absorption/extinction probes. 

Then the observed Ly$\alpha$ line luminosity writes
\begin{equation}\label{eq:Llyaobs}
L_{{\mathrm{Ly}}\alpha}^{\mathrm{obs}}\simeq 4.36 \times 10^{42}\,\left(\frac {\dot{M}_{\star}}{M_{\odot}\,\mathrm{yr}^{-1}}\right)\,f^{\rm dust}_{912}\,(1-f^{\mathrm{HI}}_{912}) \,f^{\mathrm{dust}}_{{\mathrm{Ly}}\alpha}\,f^{\mathrm{IGM}}_{{\mathrm{Ly}}\alpha}~\mathrm{erg\,s}^{-1}.
\end{equation}
{On the whole, this equation contains four parameters: the three parameters in the definition of $\tau_{\rm HI}$ (Equation~(\ref{eq:tauHI})), i.e., $\tau^0_{\rm HI}$, $\alpha_{\rm HI}$, and $\beta_{\rm HI}$, plus $\gamma$ (Equation~(\ref{eq:tau_dust_ion})). We have attempted to determine their values fitting the Ly$\alpha$ line LFs by \citet{Blanc2011} at $z \sim 3$ and by \citet{Ouchi2008} at $z \sim 3.8$. However, we could not find an unambiguous solution because of the strong degeneracy among the parameters. To break the parameters' degeneracy, we fixed $\alpha_{\rm HI}= 0.25$, in analogy to Equation~(\ref{eq|extgigi}), and $\tau^0_{\rm HI} = 0.3$. Furthermore, we discarded the solutions that imply too high emission rates of ionizing photons and too low electron scattering optical depth (see below). Based on these, somewhat loose, criteria, we have chosen $\beta_{\rm HI}\simeq 1.5$ and $\gamma\simeq 0.85$. 	} 	

{A check on the validity of our choices is provided by recent Lyman-continuum imaging of galaxies at $z\simeq 3$.} \citet{Nestor2013} measured the average ratios of non-ionizing UV to Lyman-continuum flux density corrected for IGM attenuation, $\eta_{\rm obs}$, for a sample  of LBGs and a sample of LAEs, both at $z\sim 3$. Such ratios were found to be $\eta_{\rm LBG}=18.0^{+34.8}_{-7.4}$ for LBGs (rest-frame UV absolute magnitudes $-22\leqslant M_{\rm UV} \leqslant -20$) and $\eta_{\rm LAE}=3.7^{+2.5}_{-1.1}$ for LAEs ($-20< M_{\rm UV} \leqslant -18.3$). {\citet{Mostardi2013} probed the Lyman-continuum spectral region of 49 LBGs and 70 LAEs spectroscopically confirmed at $z \sim 2.85$, as well as of 58 LAE photometric candidates in the same redshift range. After correcting for foreground galaxy contamination and HI absorption in the IGM, the average values for their samples are $\eta_{\rm LBG}=82\pm 45$, $\eta_{\rm LAE}=7.6\pm 4.1$ for the spectroscopic sample and $\eta_{\rm LAE}=2.6\pm 0.8$ for the full LAE sample. }

The observed ratios are equal to the intrinsic ones $\eta_{\rm int} \simeq R_{\rm int}^{-1}$ attenuated by dust and HI absorption (the latter only for ionizing photons)
\begin{equation}\label{eq:eta}
	 \eta_{\rm obs}=\eta_{\rm int}{f^{\rm dust}_{1350}\over f^{\rm dust}_{912}\cdot f^{\mathrm{HI}}_{912}}.
\end{equation}
%
%
%
%
%
The intrinsic ratio we have adopted, $\eta_{\rm int} \simeq R_{\rm int}^{-1}\simeq 3.77$, is within the range measured for LAEs in both the \citet{Nestor2013} and the \citet{Mostardi2013} samples, implying that the attenuation both by dust and by HI is small, as expected in the present framework (cf. Figure~\ref{fig:properties_LBGs_LAEs}). {LBGs in both samples have SFRs in the range 5--$250\,M_\odot\,\pyr$, with median values around $50\,M_\odot\,\pyr$, and gas metallicities $Z_{\rm g}\simeq 0.7 \pm 0.3\,Z_\odot$. After Equation~(\ref{eq:eta}) the optical depth for absorption of ionizing photons by HI is $\tau_{\rm HI,LBG} = \ln(\eta_{\rm obs,LBG}/\eta_{\rm int}) - \ln(f^{\rm dust}_{1350}/f^{\rm dust}_{912}) = \ln(\eta_{\rm obs,LBG}/\eta_{\rm int}) - (\gamma - 1) A_{1350} / 1.08$,
giving $\tau_{\rm HI,LBG} \simeq 1.8^{+1.3}_{-0.6} $ for $\eta_{\rm obs,LBG} = 18.0^{+34.8}_{-7.4}$  \citep{Nestor2013} and $\tau_{\rm HI,LBG} \simeq 3.3^{+0.6}_{-0.9} $ for $\eta_{\rm obs,LBG} = 82 \pm 45$ \citep{Mostardi2013}. With our choice for the parameters, Equation~(\ref{eq:tauHI}) gives $\tau_{\rm HI} = \tau^0_{\rm HI} \dot M_\star^{\alpha_{\rm HI}} + \beta_{\rm HI} \simeq 2.3^{+0.4}_{-0.4}$, consistent with both observational estimates}.

As illustrated by Figure~\ref{fig:LF_LAEs_c}, the Ly$\alpha$ line LFs yielded by the model compare quite well with observational determinations at several redshifts, from $z=3$ to $z=7.7$. The bottom-right panel shows that the \textit{intrinsic} evolution of the Ly$\alpha$ line LF is remarkably weak from $z=2$ to $z=6$, even weaker than in the UV case (Figure~\ref{fig:LF_LBGs_1350}), and similarly to the case of the UV LF, its faint portion is predicted to steepen with increasing redshift. The \textit{observed} evolution at high-$z$ is more strongly negative than the \textit{intrinsic} one due to the increasing attenuation by the IGM. The figure also demonstrates that, at $z\geqslant 5.7$, observational estimates based on photometric samples (open symbols), only partially confirmed in spectroscopy, tend to be systematically higher than those based on purely spectroscopic samples (filled symbols). Therefore, more extensive spectroscopic confirmation is necessary before firm conclusions on the high-$z$ evolution of the Ly$\alpha$ line LF can be drawn.

The average properties, weighted by the halo formation rate, of LAEs at $z=2$, 6, and 12 are shown, as a function of the observed Ly$\alpha$ luminosity, in the right-hand panels of Figure~\ref{fig:properties_LBGs_LAEs}. Compared to LBGs (left panels of the same figure), LAEs have somewhat younger ages, implying somewhat lower stellar masses, SFRs, and metallicities at given $M_{\rm vir}$. The latter two factors combine to give substantially lower dust extinction.

\subsection{Escape fraction of ionizing photons and reionization}\label{sect:esc}

The injection rate of ionizing photons into the IGM is
\begin{equation}\label{eq:Emission_rate}
\dot N^{\rm esc}_{912}=\dot N^{\rm int}_{912} f^{\rm esc}_{912} =  k_{\rm ion} {\dot M_\star}  f^{\rm esc}_{912} \ \hbox{photons}\,\hbox{s}^{-1},
\end{equation}
where $f^{\rm esc}_{912} \equiv f^{\rm dust}_{912} \times f^{\rm HI}_{912}$ is the fraction of ionizing photons emerging at the galaxy boundary. The dependencies on UV magnitude and redshift of $f^{\rm dust}_{912}$, $f^{\rm HI}_{912}$, and $f^{\rm esc}_{912}$, weighted by the halo formation rate, are illustrated in Figure~\ref{fig:fesc_Mz2}.

{As shown by Equations~(\ref{eq|extgigi}) and (\ref{eq:tauHI}), the optical depths for absorption by both dust and HI (and thus the corresponding escape fractions) are determined by the intrinsic properties of the galaxies (the SFR and, in the case of dust absorption, the metallicity), mostly controlled by the halo mass. They are thus weakly dependent on redshift. Lower mass galaxies have lower SFRs and, more importantly, lower metallicities. This translates into substantially lower optical depths ($<1$), i.e., substantially higher escape fractions. For brighter galaxies, which have optical depths $\geqslant 1$, the escape fractions are exponentially sensitive to the weak redshift dependence of the metallicity at given halo mass (see Figure~\ref{fig:sph_evol_oqs1}). Thus the small decrease with $z$ of the metallicity  translates, for bright galaxies, into a significant increase of $f^{\rm dust}_{912}$. These luminosity and redshift dependencies do not apply to $f^{\rm HI}_{912}$ which, not being affected by metallicity, is almost redshift independent at all luminosities.  In the bottom-right panel of Figure~\ref{fig:fesc_Mz2} the escape fractions of ionizing photons given by the model for two ranges of observed UV luminosity are compared with observational estimates at several redshifts. The agreement is generally good. }

The redshift-dependent emission rate function, $\phi(\log \dot N^{\rm esc}_{912}, z)$, can be constructed in the same way as the UV LF (Section~\ref{sect:UV_LFs}). The average injection rate of ionizing photons into the IGM per unit comoving volume at redshift $z$ is
\begin{equation}
	\langle \dot N^{\rm esc}_{912} \rangle (z) = \int \dot N^{\rm esc}_{912} \phi(\log \dot N^{\rm esc}_{912}, z) d\log \dot N^{\rm esc}_{912}.
\end{equation}
{Figure~\ref{fig:dN912dtdV} compares the average injection rate of ionizing photons into the IGM, $\langle\dot N^{\rm esc}_{912}\rangle(z)$, as a function of redshift yielded by the model for two choices of the critical halo mass ($M_{\rm crit}=10^{8.5}\,M_\odot$ and $M_{\rm crit}=10^{10}\,M_\odot$) with observational estimates. The original data refer to different quantities such as the proper hydrogen photoionization rate, $\Gamma_{\rm HI}(z)$, the average specific intensity of UV background, $J_\nu(z)$, and the comoving spatially averaged emissivity, $\epsilon_\nu(z)$. The conversion of these quantities into $\langle\dot N^{\rm esc}_{912}\rangle(z)$ was done using the formalism laid down by \citet{KuhlenFaucherGiguere2012}. The values of $\langle\dot N^{\rm esc}_{912}\rangle$ obtained from the model are above the estimates by \citet{KuhlenFaucherGiguere2012} but consistent with (although on the high side of) the more recent results by \citet{BeckerBolton2013} and \citet{Nestor2013}.
}

The reionization of the IGM is described in terms of the evolution of the volume filling factor of HII regions, $Q_{\rm HII}(t)$, which is ruled by \citep{Madau1999}
\begin{equation}\label{eq:Q_HII}
	\frac{dQ_{\rm HII}}{dt} \simeq \frac{\langle\dot N^{\rm esc}_{912}\rangle}{\bar n_{\rm H}} - \frac{Q_{\rm HII}}{\bar t_{\rm rec}},
\end{equation}
where $\bar n_{\rm H} = X \rho_{\rm c} \Omega_{\rm b} / m_{\rm p} \simeq 2.5 \times 10^{-7} X (\Omega_{\rm b} h^2/0.022)$ $\hbox{cm}^{-3}$ is the mean comoving hydrogen number density in terms of the primordial mass fraction of hydrogen $X = 0.75$, of the present-day critical density $\rho_{\rm c} = 1.878 h^2 \times 10^{-29}\ \gpcmcube$, and of the proton mass $m_{\rm p} = 1.67 \times 10^{-24}\ \gram$. The mean recombination time is given by \citep{Madau1999,KuhlenFaucherGiguere2012}
\begin{eqnarray}
	\bar t_{\rm rec} & = & \frac{1}{f_e \bar n_{\rm H} (1+z)^3 \alpha_{\rm B}(T_0) C_{\rm HII}} = \nonumber \\
&=& {0.51\over f_e} \Big( \frac{\Omega_{\rm b}}{0.049} \Big)^{-1} \Big( \frac{1+z}{7} \Big)^{-3} \Big( \frac{T_0}{2 \times 10^4\ {\rm K}} \Big)^{0.7} \Big( \frac{C_{\rm HII}}{6} \Big)^{-1}\ \Gyr,
\end{eqnarray}
where $f_e$ is the number of free electrons per hydrogen nucleus in the ionized IGM, assumed to have a temperature $T_0 = 2 \times 10^4\ \rm K$, and $C_{\rm HII} \equiv \langle \rho^2_{\rm HII} \rangle / \langle \rho_{\rm HII} \rangle^2$ is the clumping factor of the ionized hydrogen. $f_e$ depends on the ionization state of helium; we assumed it to be doubly ionized ($f_e = 1 + Y/2X \simeq 1.167$) at $z < 4$ and singly ionized ($f_e = 1 + Y/4X \simeq 1.083$) at higher redshifts \citep{Robertson2013}.

{The clumping factor has been extensively investigated using numerical simulations \citep[see][for a critical discussion of earlier work]{Finlator2012}. A series of drawbacks have been progressively discovered and corrected. The latest studies, accounting for the photo-ionization heating, that tends to smooth the diffuse IGM, and for the IGM temperature, which could suppress the recombination rate further, generally find relatively low values of $C_{\rm HII}$ \citep[][see the upper left panel of Figure~\ref{fig:Reion_tau}]{Pawlik2009,McQuinn2011,Shull2012,Finlator2012,KuhlenFaucherGiguere2012}. Alternatively, the clumping factor can be computed as the second moment of the IGM density distribution, integrating up to a maximum overdensity \citep{Kulkarni2013}. We adopt, as our reference, the model $C_{\rm HII,T_b, x_{\rm HII}>0.95}$ by \citet{Finlator2012}, but we will discuss the effect of different choices. }

Additional constraints on the reionization history are set by the electron scattering optical depth, $\tau_{\rm es}$, measured by Cosmic Microwave Background (CMB) anisotropy experiments. The optical depth of electron scattering up to redshift $z$ is given by
\begin{equation}
	\tau_{\rm es}(\leqslant z) = \int^z_0 dz' \Big| \frac{dt}{dz'} \Big| c \sigma_{\rm T} n_e(z') = \int^z_0 dz' \frac{c (1+z')^2}{H_0 \sqrt{\Omega_{\Lambda} + \Omega_{\rm m} (1+z')^3}} f_e Q_{\rm HII}(z') \sigma_{\rm T} \bar n_{\rm H},
\end{equation}
where $n_e \simeq f_e Q_{\rm HII} \bar n_{\rm H} (1+z)^3$ is the mean electron density, $c$ is the speed of light, and $\sigma_{\rm T} \simeq 6.65 \times 10^{-25}\ \cmsquare$ is the Thomson cross section. WMAP 9-yr data alone give $\tau_{\rm es}=0.089\pm 0.014$, slightly decreasing to $\tau_{\rm es}=0.081\pm 0.012$ when they are combined with external data sets \citep{Hinshaw2013}. The combination of the \textit{Planck} temperature power spectrum with a WMAP polarisation low-multipole likelihood results in an estimate closely matching the WMAP 9-yr value, $\tau_{\rm es}=0.089^{+0.012}_{-0.014}$ \citep{PlanckCollaborationXVI2013}. However, replacing the WMAP polarised dust template with the far more sensitive \textit{Planck}/HFI 353\,GHz polarisation map lowers the best fit value to $\tau_{\rm es}=0.075\pm 0.013$ \citep{PlanckCollaborationXV2013}; this result has however to be taken as preliminary.

The evolution of $Q_{\rm HII}(t)$ and of the electron scattering optical depth $\tau_{\rm es}(\leqslant z)$ yielded by the model adopting the critical halo mass $M_{\rm crit} = 10^{8.5}\ M_\odot$, the effective escape fraction $f^{\rm esc}_{912}$ laid down before, and the evolutionary law for the IGM clumping factor $C_{\rm HII}$ by {\citet{Finlator2012}, are shown by the thick solid black lines in the main panel of Figure~\ref{fig:Reion_tau}. Although this model gives a $\tau_{\rm es}$ consistent with the determination by \citet{PlanckCollaborationXVI2013} and less than $2\sigma$ below those by \citet{Hinshaw2013} and \citet{PlanckCollaborationXVI2013},} it yields a more extended fully ionized phase than indicated by the constraints on $Q_{\rm HII}$ at $z\geqslant 6$ collected by \citet{Robertson2013} who, however, cautioned that they are all subject to substantial systematic or modeling uncertainties. {Indications pointing to a rapid drop of the ionization degree above $z\simeq 6$ include hints of a decrease of the comoving emission rates of ionizing photons (see Figure~\ref{fig:dN912dtdV}), of sizes of quasar near zones, and of the Ly$\alpha$ transmission through the IGM \citep[see][for references]{Robertson2013}.}

As we have seen before, the observed UV LFs imply $M_{\rm crit}\lesssim 10^{10}\,M_\odot$. Adopting the latter value and keeping our baseline $f^{\rm esc}_{912}(z)$ and $C_{\rm HII}(z)$ shortens the fully ionized phase that now extends only up to $z \sim 6$ (dotted red line in the main figure of Figure~\ref{fig:Reion_tau}), lessening the discrepancy with constraints on $Q_{\rm HII}$ at the cost of  $\tau_{\rm es}$ dropping almost $\simeq 3\,\sigma$ below the best fit WMAP value and $2\,\sigma$ below the best fit value of \citet{PlanckCollaborationXV2013}. This conclusion is unaffected by different choices for $C_{\rm HII}(z)$, illustrated in the upper left panel of Figure~\ref{fig:Reion_tau}, as long as we keep our baseline $f^{\rm esc}_{912}(z)$. This is because the production rate of ionizing photons (first term on the right-hand side of Equation~(\ref{eq:Q_HII})) is always substantially larger than the recombination rate. 

The minimum SFR density required to keep the universe fully ionized at the redshift $z$ is \citep{Madau1999}
\begin{equation}\label{eq:SFR_min}
	\dot \rho_\star \simeq 8.2 \times 10^{-4} \Big( \frac{C_{\rm HII}}{f_{\rm esc}} \Big) \Big( \frac{\Omega_{\rm b} h^2}{0.022} \Big)^2 \Big( \frac{1+z}{7} \Big)^3\ M_\odot\ {\rm yr}^{-1}\ {\rm Mpc}^{-3}.
\end{equation}
It is shown in Figure~\ref{fig:SFRD_8d5_18d0} (grey area) for $3 \lesssim C_{\rm HII}/f_{\rm esc} \lesssim 30$. {A similar figure was presented by \citet[][their Figure~3]{Finkelstein2012}. A comparison of their green curve with our blue curve illustrates the difference between the UV luminosity density implied by our model for sources brighter than $M_{\rm UV}=-18$ and that from the hydrodynamic simulations of \citet{Finlator2011a}.}

{A sort of tradeoff between the constraints on $Q_{\rm HII}$ and those on $\tau_{\rm es}$ is obtained if the cooling rate of HII increases rapidly with decreasing $z$ for $z\lesssim 8$. This might be the case if, for example, the clumping factor climbs in this redshift range, as in the $C_{-1}$ L6N256 no-reheating simulation by \citet[][dot-dashed blue line in the main figure of Figure~\ref{fig:Reion_tau}]{Pawlik2009}. The drawbacks suffered by these no-reheating simulations were, however, pointed out by \citet{Pawlik2009} and \citet{Finlator2012}. }

{The tension between the determinations of $\tau_{\rm es}(\leqslant z)$ and constraints on $Q_{\rm HII}(z)$ has been repeatedly  discussed in recent years \citep[e.g.,][]{KuhlenFaucherGiguere2012,HaardtMadau2012,Robertson2013}.  Our conclusions are broadly consistent with the earlier ones, as can be seen by all models matching the observed high-$z$ UV LFs. However, our model differs from the others because our LFs and the escape fraction as a function of redshift are physically grounded, while the quoted models are based on phenomenological fits to the data. This means that our model is more constrained; for example, the extrapolations of the LFs outside the luminosity and redshift ranges covered by observations come out from our equations rather than being controlled by adjustable parameters. In other words, we explore different parameter spaces. As a result, we differ in important details such as the slope of the faint tail of the UV LF and its redshift dependence, the total number density of high-$z$ UV galaxies, and the redshift dependence of the escape fraction of ionizing photons. }

{The need for a redshift- or mass dependence of $f^{\rm esc}_{912}$ has also been deduced by other studies \citep[e.g.,][]{Alvarez2012,Mitra2013}. \citet{Mitra2013} combined data-constrained reionization histories and the evolution of the LF of early galaxies to find an empirical indication of a 2.6 times increase of the average escape fraction from $z = 6$ to $z = 8$. \citet{Alvarez2012} argued that there are both theoretical and observations indications that $f^{\rm esc}_{912}$ is higher at lower halo masses and proposed that a faint population of galaxies with host halo masses of $\sim 10^{8-9}\,M_\odot$ dominated the ionizing photon budget at redshifts $z\gtrsim 9$ due to their much higher escape fractions, again empirically estimated. Our model provides a physical basis for the increase of the effective $f^{\rm esc}_{912}$ with mass and redshift.}

The present data do not allow us to draw any firm conclusion on the reionization history because they may be affected by substantial uncertainties. Those uncertainties on data at $z=6$--7 are discussed by \citet{Robertson2013}. Those on $\tau_{\rm es}$ are illustrated by the finding \citep{PlanckCollaborationXV2013} that different corrections for the contamination by polarised foregrounds may lower the best fit value by about $1\,\sigma$.

\section{Conclusions}\label{sect:conclusions}

We have worked out a physical model for the evolution of the UV LF of high-$z$ galaxies and for the reionization history. The LF is directly linked to the formation rate of virialized halos and to the cooling and heating processes governing the star formation. For the low halo masses and young galactic ages of interest here it is not enough to take into account SN and AGN feedback, as usually done for halo masses $M_{\rm vir}\gtrsim 10^{11}\,M_\odot$, because other heating processes, such as the radiation from massive low-metallicity stars, stellar winds, and the UV background, can contribute to reducing and eventually quenching the SFR. We have modeled this by increasing the efficiency of cold gas removal and introducing a lower limit, $M_{\rm crit}$, to halo masses that can host active star formation.

Another still open issue is the production rate of UV photons per unit halo mass at high-$z$, which is influenced by two competing effects. On one side, the expected increase with redshift of the Jeans mass, hence of the characteristic stellar mass, entails a higher efficiency in the production of UV photons. On the other side, more UV photons imply more gas heating, i.e., a decrease of the SFR. We find that the observed UV LFs up to the highest redshifts are very well reproduced with the SFRs yielded by the model and the extinction law of Equation~(\ref{eq|extgigi}) for a production rate of UV photons corresponding to a \citet{Chabrier2003} IMF.

The observed UV LFs (Figure~\ref{fig:LF_LBGs_1350}) constrain $M_{\rm crit}$ to be $\lesssim 10^{10}\,M_\odot$, consistent with estimates from simulations. Figure~\ref{fig:LF_LBGs_1350} highlights several features of the model: i) dust extinction is higher for higher luminosities, associated to more massive halos which have a faster metal enrichment; ii) the higher feedback efficiency in less massive halos makes the slope of the faint end of the LF flatter than that of the halo formation rate; yet the former reflects to some extent the steepening with increasing $z$ of the latter; this has important implications for the sources of the ionizing background at high $z$; iii) the evolution of the LF from $z=2$ to $z=6$ is weak because the decrease with increasing redshift of the halo formation rate in the relevant range of halo masses is largely compensated by the increase of the star formation efficiency due to the faster gas cooling {and by the increase of dust extinction with increasing halo mass}.

\begin{figure*} 
\begin{center}
	\includegraphics[width=0.7\textwidth]{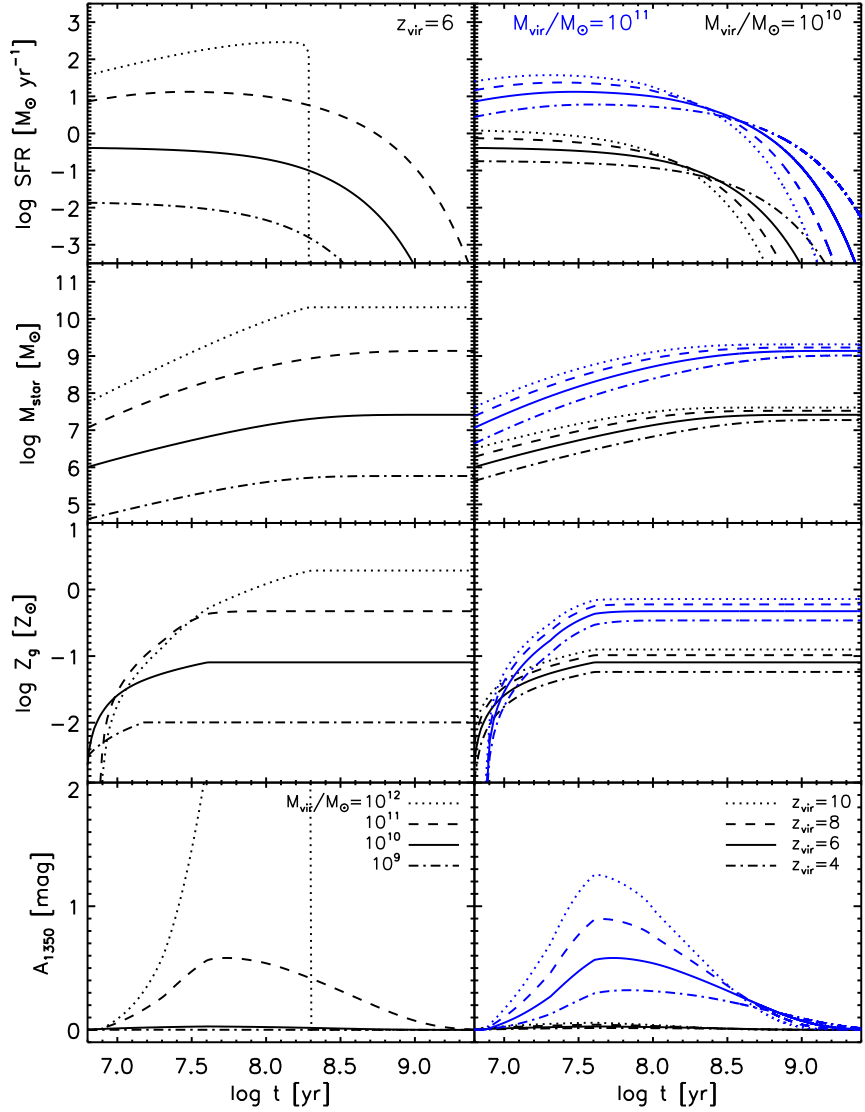}
	\caption{\textit{Left panels\/}: Evolution with \textit{galactic age} of the SFR, of the stellar mass, $M_{\rm star}$, of the gas metallicity, $Z_{\rm g}$, and of the dust extinction, $A_{1350}$ (from top to bottom) for halo masses of  $M_{\rm vir} = 10^{9}$ (dot-dashed lines), $10^{10}$ (solid lines), $10^{11}$ (dashed lines), and $10^{12}\ M_\odot$ (dotted lines) virialized at $z_{\rm vir} = 6$. {The galactic age is measured from the virialization redshift, i.e., $t=0$ for $z=z_{\rm vir}$.} \textit{Right panels\/}: Evolution of the quantities on the left panels at fixed halo mass ($M_{\rm vir} = 10^{10}$ and $10^{11}\ M_\odot$, black and blue lines, respectively) for different redshifts: $z_{\rm vir} = 4$ (dot-dashed lines), 6 (solid lines), 8 (dashed lines), and 10 (dotted lines).
}\label{fig:sph_evol_oqs1}
\end{center}
\end{figure*}

\begin{figure} 
\centering
	\includegraphics[width=0.8\textwidth]{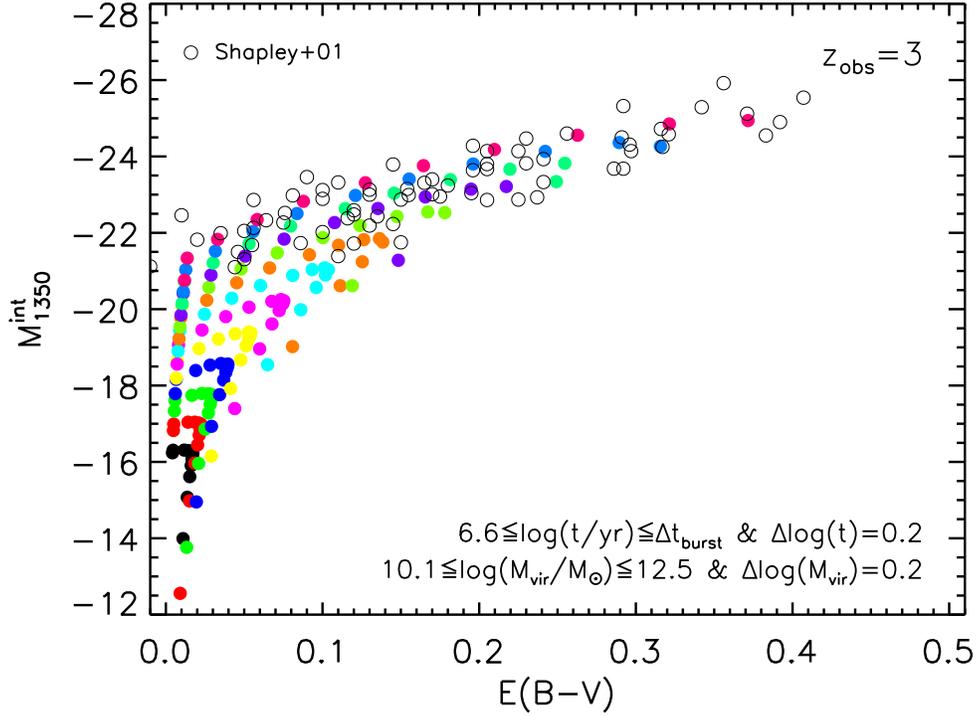}
\caption{{Correlation of the intrinsic (extinction-corrected) rest-frame absolute UV magnitude, $M_{1350}^{\mathrm{int}}$, with the color excess, $E(B-V)$. The open circles are the data by \citet[][cf. their Figure 11,]{Shapley2001} corrected for the different cosmology used here, the filled circles show the expectations of our model using the extinction law of Equation~(\ref{eq|extgigi}), for halo masses in the range $10^{10}\, M_{\odot}\la M_{\rm vir} \la 4\times 10^{12}\, M_{\odot}$ sampled in intervals $\Delta \log M_{\rm vir} = 0.2$, and for ages in the range $4\times 10^6\la t/{\rm yr} \la \Delta t_{\mathrm{burst}}(M_{\rm vir}, z_{\rm obs})$ (see Appendix of \citet{Fan2010} for an approximation of the duration of the star formation phase, $\Delta t_{\mathrm{burst}}$, as a function of halo mass and redshift). The relation $E(B-V)\approx A_{1350}/11$ by \citet{Calzetti2000} has been adopted.}
}\label{fig:M_EBV}
\end{figure}

Another key property of the model (Figure~\ref{fig:sph_evol_oqs1}) is the fast metal enrichment of the more massive galaxies that translates into a rapid increase of dust obscuration. Therefore these galaxies show up mostly at far-IR/(sub-)millimeter wavelengths, a prediction successfully tested against observational data (Figures~\ref{fig:SFRF_z} and \ref{fig:SFRD_8d5_18d0}). {The model thus predicts a strong correlation between dust attenuation and halo/stellar mass for UV selected high-$z$ galaxies.} The ratio of dust-obscured to unobscured star formation has a broad maximum at $z\simeq 2$--3. The decrease at lower redshifts is due to the decreasing amount of ISM in galaxies; at higher redshifts it is related to the fast decrease of the abundance of massive halos where the metal enrichment and, correspondingly, the dust extinction grow fast.

Similarly, good fits are obtained for the Ly$\alpha$ line LFs (Figure~\ref{fig:LF_LAEs_c}) that provide information on the production rate of ionizing photons and on their absorption by neutral interstellar hydrogen. Further constraints on the attenuation by dust and HI are provided by recent  measurements \citep{Nestor2013,Mostardi2013} of the observed ratios of non-ionizing UV to Lyman-continuum flux densities for LAEs and LBGs. These data have allowed us to derive a simple relationship between the optical depth for HI absorption and SFR. Taking this relation into account, the model reproduces the very weak evolution of the Ly$\alpha$ line LF between $z=2$ and $z=6$, even weaker than in the UV.

{The derived relationships linking the optical depths for absorption of ionizing photons by dust and HI to the SFR and, in the case of dust absorption, to the metallicity of the galaxies, imply higher \textit{effective} escape fractions for galaxies with lower intrinsic UV luminosities or lower halo/stellar masses, and also a mild increase of the escape fraction with increasing redshift at fixed luminosity or halo/stellar mass. Redshift- or mass-dependencies of the escape fraction were previously empirically deduced by, e.g., \citet{Alvarez2012} and \citet{Mitra2013}. Our model provides a physical basis for these dependencies.}

At this point we can compute the average injection rate of ionizing photons into the IGM as a function of halo mass and redshift. To reconstruct the ionization history of the universe we further need the evolution of the clumping factor of the IGM, {for which we have adopted, as our reference, the model $C_{\rm HII,T_b, x_{\rm HII}>0.95}$ by \citet{Finlator2012}, but also considered alternative models, discussed in the literature. With our recipe for the escape fraction of ionizing photons we find that galaxies already represented in the observed UV LFs, i.e., with $M_{\rm UV}\lesssim -18$, hosted by halo masses  $\gtrsim 10^{10}\,M_\odot$, can account for a complete ionization of the IGM up to $z\simeq 6$. To get complete ionization up to $z\simeq 7$ the population of star-forming galaxies at this redshift must extend in luminosity to $M_{\rm UV}\sim -13$ or fainter, in agreement with the conclusions of other analyses \citep[e.g.,][]{Robertson2013}. The surface densities of $M_{\rm UV}\sim -13$ galaxies would correspond to those of halo masses of $\sim 10^{8.5}\,M_\odot$, not far from the lower limit on $M_{\rm crit}$ from hydrodynamical simulations.}

A complete IGM ionization up to $z\simeq 7$ is disfavoured by some (admittedly uncertain) data at $z\simeq 6$--7 collected by \citet{Robertson2013}, that point to a fast decline of the ionization degree at $z\gtrsim 6$. {However, an even more extended ionized phase is implied by the determinations of electron scattering optical depths, $\tau_{\rm es}$, from CMB experiments. Our model adopting the critical halo mass $M_{\rm crit} = 10^{8.5}\ M_\odot$, yielding complete ionization up to $z\simeq 7$, gives a $\tau_{\rm es}$ consistent with determination by \citet{PlanckCollaborationXVI2013} and less than $2\sigma$ below those by \citet{Hinshaw2013} and \citet{PlanckCollaborationXVI2013}. Raising $M_{\rm crit}$ to $10^{10}\ M_\odot$ limits the fully ionized phase to $z \lesssim 6$ and decreases $\tau_{\rm es}$ to a value almost $\simeq 3\,\sigma$ below the estimates by \citet{Hinshaw2013} and \citet{PlanckCollaborationXVI2013} and $2\,\sigma$ below that by \citet{PlanckCollaborationXV2013}. } Since all these constraints on the reionization history are affected by substantial uncertainties, any firm conclusion is premature. Better data are needed to resolve the issue.

\begin{acknowledgements}
{We are grateful to the referee for a careful reading of the manuscript and many constructive comments that helped us substantially improving the paper. We also acknowledge useful comments from G. Zamorani.} Z.Y.C. acknowledges support from the joint PhD project between XMU and SISSA. The work has been supported in part by ASI/INAF agreement n. I/072/09/0 and by PRIN 2009 ``Millimeter and sub-millimeter spectroscopy for high resolution studies of primeval galaxies and clusters of galaxies''.
\end{acknowledgements}


\begin{thebibliography}{}

\bibitem[Adelberger \& Steidel(2000)]{AdelbergerSteidel2000} Adelberger, K.~L., \& Steidel, C.~C.\ 2000, \apj, 544, 218



\bibitem[Alavi et al.(2014)]{Alavi2014} Alavi, A., Siana, B., Richard, J., et al.\ 2014, \apj, 780, 143

\bibitem[Alvarez et al.(2012)]{Alvarez2012} Alvarez, M.~A., Finlator, K., \& Trenti, M.\ 2012, \apjl, 759, L38

\bibitem[Asplund et al.(2009)]{Asplund2009} Asplund, M., Grevesse, N., Sauval, A.~J., \& Scott, P.\ 2009, \araa, 47, 481

\bibitem[Becker \& Bolton(2013)]{BeckerBolton2013} Becker, G.~D., \& Bolton, J.~S.\ 2013, \mnras, 436, 1023

\bibitem[Blanc et al.(2011)]{Blanc2011} Blanc, G.~A., Adams, J.~J., Gebhardt, K., et al.\ 2011, \apj, 736, 31

\bibitem[Bouch{\'e} et al.(2013)]{Bouche2013} Bouch{\'e}, N., Murphy, M.~T., Kacprzak, G.~G., et al.\ 2013, Science, 341, 50

\bibitem[Bouwens et al.(2006)]{Bouwens2006} Bouwens, R.~J., Illingworth, G.~D., Blakeslee, J.~P., \& Franx, M.\ 2006, \apj, 653, 53

\bibitem[Bouwens et al.(2007)]{Bouwens2007} Bouwens, R.~J., Illingworth, G.~D., Franx, M., \& Ford, H.\ 2007, \apj, 670, 928

\bibitem[Bouwens et al.(2008)]{Bouwens2008} Bouwens, R.~J., Illingworth, G.~D., Franx, M., \& Ford, H.\ 2008, \apj, 686, 230

\bibitem[Bouwens et al.(2009)]{Bouwens2009} Bouwens, R.~J., Illingworth, G.~D., Franx, M., et al.\ 2009, \apj, 705, 936

\bibitem[Bouwens et al.(2011a)]{Bouwens2011a} Bouwens, R.~J., Illingworth, G.~D., Labbe, I., et al.\ 2011, \nat, 469, 504

\bibitem[Bouwens et al.(2011b)]{Bouwens2011b} Bouwens, R.~J., Illingworth, G.~D., Oesch, P.~A., et al.\ 2011, \apj, 737, 90

\bibitem[Bouwens et al.(2012)]{Bouwens2012a} Bouwens, R.~J., Bradley, L., Zitrin, A., et al.\ 2012, arXiv:1211.2230

\bibitem[Bouwens et al.(2012)]{Bouwens2012b} Bouwens, R.~J., Illingworth, G.~D., Oesch, P.~A., et al.\ 2012, \apj, 754, 83

\bibitem[Bradley et al.(2012)]{Bradley2012} Bradley, L.~D., Trenti, M., Oesch, P.~A., et al.\ 2012, \apj, 760, 108

\bibitem[Cai et al.(2013)]{Cai2013} Cai, Z.-Y., Lapi, A., Xia, J.-Q., et al.\ 2013, \apj, 768, 21

\bibitem[Calverley et al.(2011)]{Calverley2011} Calverley, A.~P., Becker, G.~D., Haehnelt, M.~G., \& Bolton, J.~S.\ 2011, \mnras, 412, 2543

\bibitem[Calzetti et al.(2000)]{Calzetti2000} Calzetti, D., Armus, L., Bohlin, R.~C., et al.\ 2000, \apj, 533, 682

\bibitem[Caputi et al.(2007)]{Caputi2007} Caputi, K.~I., Lagache, G., Yan, L., et al.\ 2007, \apj, 660, 97

\bibitem[Castellano et al.(2012)]{Castellano2012} Castellano, M., Fontana, A., Grazian, A., et al.\ 2012, \aap, 540, A39

\bibitem[Chabrier(2003)]{Chabrier2003} Chabrier, G.\ 2003, \pasp, 115, 763



\bibitem[Coe et al.(2013)]{Coe2013} Coe, D., Zitrin, A., Carrasco, M., et al.\ 2013, \apj, 762, 32

\bibitem[Cucciati et al.(2012)]{Cucciati2012} Cucciati, O., Tresse, L., Ilbert, O., et al.\ 2012, \aap, 539, A31

\bibitem[Cullen et al.(2013)]{Cullen2013} Cullen, F., Cirasuolo, M., McLure, R.~J., \& Dunlop, J.~S.\ 2013, arXiv:1310.0816

\bibitem[Dall'Aglio et al.(2009)]{DallAglio2009} Dall'Aglio, A., Wisotzki, L., \& Worseck, G.\ 2009, arXiv:0906.1484


\bibitem[Dav{\'e} et al.(2010)]{Dave2010} Dav{\'e}, R., Finlator, K., Oppenheimer, B.~D., et al.\ 2010, \mnras, 404, 1355

\bibitem[Dawson et al.(2007)]{Dawson2007} Dawson, S., Rhoads, J.~E., Malhotra, S., et al.\ 2007, \apj, 671, 1227

\bibitem[Dayal et al.(2013)]{Dayal2013} Dayal, P., Dunlop, J.~S., Maio, U., \& Ciardi, B.\ 2013, \mnras, 434, 1486

\bibitem[Dayal et al.(2008)]{Dayal2008} Dayal, P., Ferrara, A., \& Gallerani, S.\ 2008, \mnras, 389, 1683

\bibitem[Dayal et al.(2009)]{Dayal2009} Dayal, P., Ferrara, A., Saro, A., et al.\ 2009, \mnras, 400, 2000

\bibitem[Dekel et al.(2009)]{Dekel2009} Dekel, A., Birnboim, Y., Engel, G., et al.\ 2009, \nat, 457, 451

\bibitem[Dekel \& Birnboim(2006)]{DekelBirnboim2006} Dekel, A., \& Birnboim, Y.\ 2006, \mnras, 368, 2

\bibitem[Dijkstra et al.(2004)]{Dijkstra2004} Dijkstra, M., Haiman, Z., \& Loeb, A.\ 2004, \apj, 613, 646


\bibitem[Duval et al.(2014)]{Duval2014} Duval, F., Schaerer, D., {\"O}stlin, G., \& Laursen, P.\ 2014, \aap, 562, 52

\bibitem[Ellis et al.(2013)]{Ellis2013} Ellis, R.~S., McLure, R.~J., Dunlop, J.~S., et al.\ 2013, \apjl, 763, L7

\bibitem[Erb et al.(2006)]{Erb2006} Erb, D.~K., Shapley, A.~E., Pettini, M., et al.\ 2006, \apj, 644, 813

\bibitem[Fan et al. (2010)]{Fan2010} Fan, L., Lapi, A., Bressan, A., et al.\ 2010, \apj, 718, 1460

\bibitem[Fagotto et al.(1994a)]{Fagotto1994a} Fagotto, F., Bressan, A., Bertelli, G., \& Chiosi, C.\ 1994, \aaps, 104, 365

\bibitem[Fagotto et al.(1994b)]{Fagotto1994b} Fagotto, F., Bressan, A., Bertelli, G., \& Chiosi, C.\ 1994, \aaps, 105, 29

\bibitem[Fardal et al.(2007)]{Fardal2007} Fardal, M.~A., Katz, N., Weinberg, D.~H., \& Dav{\'e}, R.\ 2007, \mnras, 379, 985

\bibitem[Finkelstein et al.(2012)]{Finkelstein2012} Finkelstein, S.~L., Papovich, C., Ryan, R.~E., et al.\ 2012, \apj, 758, 93

\bibitem[Finlator et al.(2011a)]{Finlator2011a} Finlator, K., Dav{\'e}, R., {\"{O}}zel, F.\ 2011a, \apj, 743, 169

\bibitem[Finlator et al.(2011b)]{Finlator2011b} Finlator, K., Oppenheimer, B.~D., \& Dav{\'e}, R.\ 2011b, \mnras, 410, 1703

\bibitem[Finlator et al.(2012)]{Finlator2012} Finlator, K., Oh, S.~P., {\"O}zel, F., \& Dav{\'e}, R.\ 2012, \mnras, 427, 2464

\bibitem[Garel et al.(2012)]{Garel2012} Garel, T., Blaizot, J., Guiderdoni, B., et al.\ 2012, \mnras, 422, 310

\bibitem[Gonzalez-Perez et al.(2013)]{GonzalezPerez2013} Gonzalez-Perez, V., Lacey, C.~G., Baugh, C.~M., Frenk, C.~S., \& Wilkins, S.~M.\ 2013, \mnras, 429, 1609

\bibitem[Granato et al.(2004)]{Granato2004} Granato, G.~L., De Zotti, G., Silva, L., Bressan, A., \& Danese, L.\ 2004, \apj, 600, 580

\bibitem[Gronwall et al.(2007)]{Gronwall2007} Gronwall, C., Ciardullo, R., Hickey, T., et al.\ 2007, \apj, 667, 79

\bibitem[Gruppioni et al.(2013)]{Gruppioni2013} Gruppioni, C., Pozzi, F., Rodighiero, G., et al.\ 2013, \mnras, 432, 23

\bibitem[Haardt \& Madau(2012)]{HaardtMadau2012} Haardt, F., \& Madau, P.\ 2012, \apj, 746, 125

\bibitem[Hambrick et al.(2011)]{Hambrick2011} Hambrick, D.~C., Ostriker, J.~P., Johansson, P.~H., \& Naab, T.\ 2011, \mnras, 413, 2421

\bibitem[Hathi et al.(2008)]{Hathi2008} Hathi, N.~P., Malhotra, S., \& Rhoads, J.~E.\ 2008, \apj, 673, 686

\bibitem[Hayes et al.(2011)]{Hayes2011} Hayes, M., Schaerer, D., {\"O}stlin, G., et al.\ 2011, \apj, 730, 8

\bibitem[Heinis et al.(2014)]{Heinis2014} Heinis, S., Buat, V., Bethermin, M., et al.\ 2014, \mnras, 437, 1268

\bibitem[Hibon et al.(2010)]{Hibon2010} Hibon, P., Cuby, J.-G., Willis, J., et al.\ 2010, \aap, 515, 97

\bibitem[Hinshaw et al.(2013)]{Hinshaw2013} Hinshaw, G., Larson, D., Komatsu, E., et al.\ 2013, \apjs, 208, 19

\bibitem[Hu et al.(2010)]{Hu2010} Hu, E.~M., Cowie, L.~L., Barger, A.~J., Capak, P., Kakazu, Y., \& Trouille, L.\ 2010, \apj, 725, 394

\bibitem[Iliev et al.(2007)]{Iliev2007} Iliev, I.~T., Mellema, G., Shapiro, P.~R., \& Pen, U.-L.\ 2007, \mnras, 376, 534


\bibitem[Iwata et al.(2009)]{Iwata2009} Iwata, I., Inoue, A.~K., Matsuda, Y., et al.\ 2009, \apj, 692, 1287

\bibitem[Iwata et al.(2007)]{Iwata2007} Iwata, I., Ohta, K., Tamura, N., et al.\ 2007, \mnras, 376, 1557


\bibitem[Iye et al.(2006)]{Iye2006} Iye, M., Ota, K., Kashikawa, N., et al.\ 2006, \nat, 443, 186

\bibitem[Jaacks et al.(2012a)]{Jaacks2012a} Jaacks, J., Choi, J.-H., Nagamine, K., Thompson, R., \& Varghese, S.\ 2012a, \mnras, 420, 1606

\bibitem[Jaacks et al.(2012b)]{Jaacks2012b} Jaacks, J., Nagamine, K., \& Choi, J.~H.\ 2012b, \mnras, 427, 403


\bibitem[Jiang et al.(2013)]{Jiang2013} Jiang, L., Bian, F., Fan, X., et al.\ 2013, \apjl, 771, L6

\bibitem[Karim et al.(2011)]{Karim2011} Karim, A., Schinnerer, E., Mart{\'{\i}}nez-Sansigre, A., et al.\ 2011, \apj, 730, 61

\bibitem[Kashikawa et al.(2006)]{Kashikawa2006} Kashikawa, N., Shimasaku, K., Malkan, M.~A., et al.\ 2006, \apj, 648, 7

\bibitem[Kashikawa et al.(2011)]{Kashikawa2011} Kashikawa, N., Shimasaku, K., Matsuda, Y., et al.\ 2011, \apj, 734, 119

\bibitem[Kennicutt \& Evans(2012)]{KennicuttEvans2012} Kennicutt, R.~C., \& Evans, N.~J.\ 2012, \araa, 50, 531

\bibitem[Kere{\v s} et al.(2005)]{Keres2005} Kere{\v s}, D., Katz, N., Weinberg, D.~H., \& Dav{\'e}, R.\ 2005, \mnras, 363, 2

\bibitem[Kobayashi et al.(2010)]{Kobayashi2010} Kobayashi, M.~A.~R., Totani, T., \& Nagashima, M.\ 2010, \apj, 708, 1119

\bibitem[Kravtsov et al.(2004)]{Kravtsov2004} Kravtsov, A.~V., Gnedin, O.~Y., \& Klypin, A.~A.\ 2004, \apj, 609, 482

\bibitem[Krumholz \& Dekel(2012)]{KrumholzDekel2012} Krumholz, M.~R., \& Dekel, A.\ 2012, \apj, 753, 16

\bibitem[Kuhlen \& Faucher-Gigu{\`e}re(2012)]{KuhlenFaucherGiguere2012} Kuhlen, M., \& Faucher-Gigu{\`e}re, C.-A.\ 2012, \mnras, 423, 862

\bibitem[Kulkarni et al.(2013)]{Kulkarni2013} Kulkarni, G., Rollinde, E., Hennawi, J.~F., \& Vangioni, E.\ 2013, \apj, 772, 93

\bibitem[Lacey et al.(2011)]{Lacey2011} Lacey, C.~G., Baugh, C.~M., Frenk, C.~S., \& Benson, A.~J.\ 2011, \mnras, 412, 1828


\bibitem[Lapi \& Cavaliere(2011)]{LapiCavaliere2011} Lapi, A., \& Cavaliere, A.\ 2011, \apj, 743, 127

\bibitem[Lapi et al.(2011)]{Lapi2011} Lapi, A., Gonz{\'a}lez-Nuevo, J., Fan, L., et al.\ 2011, \apj, 742, 24

\bibitem[Lapi et al.(2013)]{Lapi2013} Lapi, A., Salucci, P., \& Danese, L.\ 2013, \apj, 772, 85

\bibitem[Lapi et al.(2006)]{Lapi2006} Lapi, A., Shankar, F., Mao, J., et al.\ 2006, \apj, 650, 42

\bibitem[Liu et al.(2008)]{Liu2008} Liu, X., Shapley, A.~E., Coil, A.~L., Brinchmann, J., \& Ma, C.-P.\ 2008, \apj, 678, 758

\bibitem[Lo Faro et al.(2009)]{LoFaro2009} Lo Faro, B., Monaco, P., Vanzella, E., et al.\ 2009, \mnras, 399, 827

\bibitem[Lorenzoni et al.(2013)]{Lorenzoni2013} Lorenzoni, S., Bunker, A.~J., Wilkins, S.~M., Caruana, J., Stanway, E.~R., \& Jarvis, M.~J.\ 2013, \mnras, 429, 150

\bibitem[Lorenzoni et al.(2011)]{Lorenzoni2011} Lorenzoni, S., Bunker, A.~J., Wilkins, S.~M., Stanway, E.~R., Jarvis, M.~J., \& Caruana, J.\ 2011, \mnras, 414, 1455

\bibitem[Madau(1995)]{Madau1995} Madau, P.\ 1995, \apj, 441, 18

\bibitem[Madau et al.(1999)]{Madau1999} Madau, P., Haardt, F., \& Rees, M.~J.\ 1999, \apj, 514, 648

\bibitem[Magnelli et al.(2011)]{Magnelli2011} Magnelli, B., Elbaz, D., Chary, R.~R., et al.\ 2011, \aap, 528, A35

\bibitem[Magnelli et al.(2013)]{Magnelli2013} Magnelli, B., Popesso, P., Berta, S., et al.\ 2013, \aap, 553, A132

\bibitem[Maiolino et al.(2008)]{Maiolino2008} Maiolino, R., Nagao, T., Grazian, A., et al.\ 2008, \aap, 488, 463

\bibitem[Malhotra \& Rhoads(2004)]{MalhotraRhoads2004} Malhotra, S., \& Rhoads, J.~E.\ 2004, \apj, 617, 5

\bibitem[Mannucci et al.(2010)]{Mannucci2010} Mannucci, F., Cresci, G., Maiolino, R., Marconi, A., \& Gnerucci, A.\ 2010, \mnras, 408, 2115

\bibitem[Mao et al.(2007)]{Mao2007} Mao, J., Lapi, A., Granato, G.~L., de Zotti, G., \& Danese, L.\ 2007, \apj, 667, 655

\bibitem[McLure et al.(2013)]{McLure2013} McLure, R.~J., Dunlop, J.~S., Bowler, R.~A.~A., et al.\ 2013, \mnras, 432, 2696

\bibitem[McLure et al.(2010)]{McLure2010} McLure, R.~J., Dunlop, J.~S., Cirasuolo, M., et al.\ 2010, \mnras, 403, 960


\bibitem[McQuinn et al.(2011)]{McQuinn2011} McQuinn, M., Oh, S.~P., \& Faucher-Gigu{\`e}re, C.-A.\ 2011, \apj, 743, 82



\bibitem[Mitra et al.(2013)]{Mitra2013} Mitra, S., Ferrara, A., \& Choudhury, T.~R.\ 2013, \mnras, 428, L1

\bibitem[Mostardi et al.(2013)]{Mostardi2013} Mostardi, R.~E., Shapley, A.~E., Nestor, D.~B., et al.\ 2013, \apj, 779, 65

\bibitem[Mu{\~n}oz(2012)]{Munoz2012} Mu{\~n}oz, J.~A.\ 2012, \jcap, 4, 15

\bibitem[Mu{\~n}oz \& Loeb(2011)]{MunozLoeb2011} Mu{\~n}oz, J.~A., \& Loeb, A.\ 2011, \apj, 729, 99

\bibitem[Murayama et al.(2007)]{Murayama2007} Murayama, T., Taniguchi, Y., Scoville, N.~Z., et al.\ 2007, \apjs, 172, 523

\bibitem[Nagamine et al.(2010)]{Nagamine2010} Nagamine, K., Ouchi, M., Springel, V., \& Hernquist, L.\ 2010, \pasj, 62, 1455

\bibitem[Navarro et al.(1997)]{Navarro1997} Navarro, J.~F., Frenk, C.~S., \& White, S.~D.~M.\ 1997, \apj, 490, 493

\bibitem[Nestor et al.(2013)]{Nestor2013} Nestor, D.~B., Shapley, A.~E., Kornei, K.~A., Steidel, C.~C., \& Siana, B.\ 2013, \apj, 765, 47

\bibitem[Oesch et al.(2012)]{Oesch2012} Oesch, P.~A., Bouwens, R.~J., Illingworth, G.~D., et al.\ 2012, \apj, 759, 135

\bibitem[Oesch et al.(2013a)]{Oesch2013a} Oesch, P.~A., Bouwens, R.~J., Illingworth, G.~D., et al.\ 2013a, \apj, 773, 75

\bibitem[Oesch et al.(2013b)]{Oesch2013b} Oesch, P.~A., Bouwens, R.~J., Illingworth, G.~D., et al.\ 2013b, arXiv:1309.2280

\bibitem[Oesch et al.(2013c)]{Oesch2013c} Oesch, P.~A., Labb{\'e}, I., Bouwens, R.~J., et al.\ 2013c, \apj, 772, 136

\bibitem[Okamoto et al.(2008)]{Okamoto2008} Okamoto, T., Gao, L., \& Theuns, T.\ 2008, \mnras, 390, 920


\bibitem[Oke \& Gunn(1983)]{OkeGunn1983} Oke, J.~B., \& Gunn, J.~E.\ 1983, \apj, 266, 713



\bibitem[Osterbrock(1989)]{Osterbrock1989} Osterbrock, D.~E.\ 1989, Astrophysics of gaseous nebulae and active galactic nuclei (Mill Valley, CA: University Science Books)

\bibitem[Ota et al.(2008)]{Ota2008} Ota, K., Iye, M., Kashikawa, N., et al.\ 2008, \apj, 677, 12

\bibitem[Ota et al.(2010)]{Ota2010} Ota, K., Iye, M., Kashikawa, N., et al.\ 2010, \apj, 722, 803

\bibitem[Ouchi et al.(2008)]{Ouchi2008} Ouchi, M., Shimasaku, K., Akiyama, M., et al.\ 2008, \apjs, 176, 301

\bibitem[Ouchi et al.(2003)]{Ouchi2003} Ouchi, M., Shimasaku, K., Furusawa, H., et al.\ 2003, \apj, 582, 60

\bibitem[Ouchi et al.(2010)]{Ouchi2010} Ouchi, M., Shimasaku, K., Furusawa, H., et al.\ 2010, \apj, 723, 869

\bibitem[Ouchi et al.(2004)]{Ouchi2004} Ouchi, M., Shimasaku, K., Okamura, S., et al.\ 2004, \apj, 611, 660

\bibitem[Padoan et al.(1997)]{Padoan1997} Padoan, P., Nordlund, A., \& Jones, B.~J.~T.\ 1997, \mnras, 288, 145

\bibitem[Pawlik et al.(2013)]{Pawlik2013} Pawlik, A.~H., Milosavljevi{\'c}, M., \& Bromm, V.\ 2013, \apj, 767, 59

\bibitem[Pawlik \& Schaye(2009)]{PawlikSchaye2009} Pawlik, A.~H., \& Schaye, J.\ 2009, \mnras, 396, L46

\bibitem[Pawlik et al.(2009)]{Pawlik2009} Pawlik, A.~H., Schaye, J., \& van Scherpenzeel, E.,\ 2009, \mnras, 394, 1812

\bibitem[Pei(1992)]{Pei1992} Pei, Y.~C.\ 1992, \apj, 395, 130


\bibitem[Planck collaboration XV(2013)]{PlanckCollaborationXV2013} Planck collaboration, Ade, P.~A.~R., Aghanim, N., et al.\ 2013, arXiv:1303.5075

\bibitem[Planck Collaboration XVI(2013)]{PlanckCollaborationXVI2013} Planck Collaboration, Ade, P.~A.~R., Aghanim, N., et al.\ 2013, arXiv:1303.5076

\bibitem[Press \& Schechter(1974)]{PressSchechter1974} Press, W.~H., \& Schechter, P.\ 1974, \apj, 187, 425

\bibitem[Rai{\v c}evi{\'c} et al.(2011)]{Raicevic2011} Rai{\v c}evi{\'c}, M., Theuns, T., \& Lacey, C.\ 2011, \mnras, 410, 775

\bibitem[Rauch et al.(2008)]{Rauch2008} Rauch, M., Haehnelt, M., Bunker, A., et al.\ 2008, \apj, 681, 856

\bibitem[Reddy \& Steidel(2009)]{ReddySteidel2009} Reddy, N.~A., \& Steidel, C.~C.\ 2009, \apj, 692, 778

\bibitem[Robertson et al.(2013)]{Robertson2013} Robertson, B.~E., Furlanetto, S.~R., Schneider, E., et al.\ 2013, \apj, 768, 71

\bibitem[Rodighiero et al.(2010)]{Rodighiero2010} Rodighiero, G., Vaccari, M., Franceschini, A., et al.\ 2010, \aap, 515, A8

\bibitem[Romano et al.(2002)]{Romano2002} Romano, D., Silva, L., Matteucci, F., \& Danese, L.\ 2002, \mnras, 334, 444

\bibitem[Salim \& Lee(2012)]{SalimLee2012} Salim, S., \& Lee, J.~C.\ 2012, \apj, 758, 134

\bibitem[Salvaterra et al.(2011)]{Salvaterra2011} Salvaterra, R., Ferrara, A., \& Dayal, P.\ 2011, \mnras, 414, 847

\bibitem[Samui et al.(2009)]{Samui2009} Samui, S., Srianand, R., \& Subramanian, K.\ 2009, \mnras, 398, 2061

\bibitem[Santos(2004)]{Santos2004} Santos, M.~R.\ 2004, \mnras, 349, 1137

\bibitem[Sasaki(1994)]{Sasaki1994} Sasaki, S.\ 1994, \pasj, 46, 427

\bibitem[Sawicki \& Thompson(2006a)]{SawickiThompson2006a} Sawicki, M., \& Thompson, D.\ 2006a, \apj, 642, 653

\bibitem[Sawicki \& Thompson(2006b)]{SawickiThompson2006b} Sawicki, M., \& Thompson, D.\ 2006b, \apj, 648, 299

\bibitem[Schenker et al.(2013)]{Schenker2013} Schenker, M.~A., Robertson, B.~E., Ellis, R.~S., et al.\ 2013, \apj, 768, 196

\bibitem[Schiminovich et al.(2005)]{Schiminovich2005} Schiminovich, D., Ilbert, O., Arnouts, S., et al.\ 2005, \apjl, 619, L47

\bibitem[Shankar et al.(2006)]{Shankar2006} Shankar, F., Lapi, A., Salucci, P., De Zotti, G., \& Danese, L.\ 2006, \apj, 643, 14

\bibitem[Shapley et al.(2001)]{Shapley2001} Shapley, A.~E., Steidel, C.~C., Adelberger, K.~L., et al.\ 2001, \apj, 562, 95

\bibitem[Sheth \& Tormen(1999)]{ShethTormen1999} Sheth, R.~K., \& Tormen, G.\ 1999, \mnras, 308, 119

\bibitem[Shibuya et al.(2012)]{Shibuya2012} Shibuya, T., Kashikawa, N., Ota, K., et al.\ 2012, \apj, 752, 114

\bibitem[Shimasaku et al.(2006)]{Shimasaku2006} Shimasaku, K., Kashikawa, N., Doi, M., et al.\ 2006, \pasj, 58, 313

\bibitem[Shull et al.(2012)]{Shull2012} Shull, J.~M., Harness, A., Trenti, M., \& Smith, B.~D.,\ 2012, \apj, 747, 100

\bibitem[Silk \& Mamon(2012)]{SilkMamon2012} Silk, J., \& Mamon, G.~A.\ 2012, Research in Astronomy and Astrophysics, 12, 917

\bibitem[Smit et al.(2012)]{Smit2012} Smit, R., Bouwens, R.~J., Franx, M., et al.\ 2012, \apj, 756, 14

\bibitem[Sobacchi \& Mesinger(2013)]{SobacchiMesinger2013} Sobacchi, E., \& Mesinger, A.\ 2013, \mnras, 432, L51

\bibitem[Sobral et al.(2009)]{Sobral2009} Sobral, D., Best, P.~N., Geach, J.~E., et al.\ 2009, \mnras, 398, 68

\bibitem[Stanway et al.(2005)]{Stanway2005} Stanway, E.~R., McMahon, R.~G., \& Bunker, A.~J.\ 2005, \mnras, 359, 1184

\bibitem[Stark et al.(2009)]{Stark2009} Stark, D.~P., Ellis, R.~S., Bunker, A., et al.\ 2009, \apj, 697, 1493

\bibitem[Stark et al.(2013)]{Stark2013} Stark, D.~P., Schenker, M.~A., Ellis, R., et al.\ 2013, \apj, 763, 129

\bibitem[Steidel et al.(1999)]{Steidel1999} Steidel, C.~C., Adelberger, K.~L., Giavalisco, M., Dickinson, M., \& Pettini, M.\ 1999, \apj, 519, 1


\bibitem[Sutherland \& Dopita(1993)]{SutherlandDopita1993} Sutherland, R.~S., \& Dopita, M.~A.\ 1993, \apjs, 88, 253

\bibitem[Tacchella et al.(2013)]{Tacchella2013} Tacchella, S., Trenti, M., \& Carollo, C.-M.\ 2013, \apjl, 768, L37


\bibitem[Tilvi et al.(2009)]{Tilvi2009} Tilvi, V., Malhotra, S., Rhoads, J.~E., et al.\ 2009, \apj, 704, 724

\bibitem[Trenti et al.(2010)]{Trenti2010} Trenti, M., Stiavelli, M., Bouwens, R.~J., et al.\ 2010, \apjl, 714, L202

\bibitem[van Breukelen et al.(2005)]{vanBreukelen2005} van Breukelen, C., Jarvis, M.~J., \& Venemans, B.~P.\ 2005, \mnras, 359, 895

\bibitem[Vanzella et al.(2010)]{Vanzella2010} Vanzella, E., Giavalisco, M., Inoue, A.~K., et al.\ 2010, \apj, 725, 1011

\bibitem[Wang et al.(2011)]{Wang2011} Wang, J., Navarro, J.~F., Frenk, C.~S., et al.\ 2011, \mnras, 413, 1373


\bibitem[Wyder et al.(2005)]{Wyder2005} Wyder, T.~K., Treyer, M.~A., Milliard, B., et al.\ 2005, \apjl, 619, L15

\bibitem[Wyithe \& Bolton(2011)]{WyitheBolton2011} Wyithe, J.~S.~B., \& Bolton, J.~S.\ 2011, \mnras, 412, 1926

\bibitem[Wyithe \& Loeb(2013)]{WyitheLoeb2013} Wyithe, J.~S.~B., \& Loeb, A.\ 2013, \mnras, 428, 2741

\bibitem[Xia et al.(2012)]{Xia2012} Xia, J.-Q., Negrello, M., Lapi, A., et al.\ 2012, \mnras, 422, 1324

\bibitem[Yan et al.(2012)]{Yan2012} Yan, H., Finkelstein, S.~L., Huang, K.~H., et al.\ 2012, \apj, 761, 177

\bibitem[Yan et al.(2011)]{Yan2011} Yan, H., Yan, L., Zamojski, M.~A., et al.\ 2011, \apj, 728, L22

\bibitem[Yoshida et al.(2006)]{Yoshida2006} Yoshida, M., Shimasaku, K., Kashikawa, N., et al.\ 2006, \apj, 653, 988

\bibitem[Zhao et al.(2003)]{Zhao2003} Zhao, D.~H., Mo, H.~J., Jing, Y.~P., B\"{o}rner, G.\ 2003, \mnras, 339, 12

\bibitem[Zheng et al.(2013)]{Zheng2013} Zheng, Z.-Y., Finkelstein, S.~L., Finkelstein, K., et al.\ 2013, \mnras, 431, 3589


\end{thebibliography}


\begin{figure} 
\centering
	\centering
	\includegraphics[width=\textwidth]{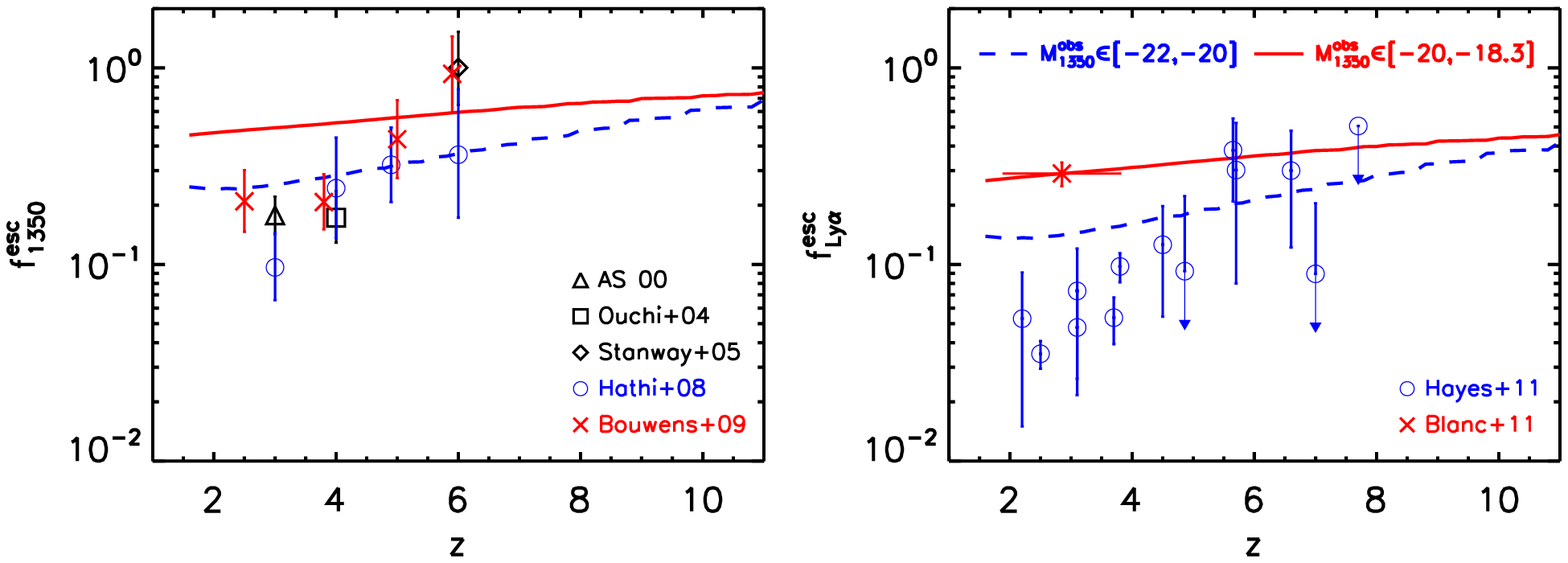}
\caption{{Average escape fractions of UV ($f^{\rm esc}_{1350}$, left panel) and Ly$\alpha$ photons ($f^{\rm esc}_{\rm Ly\alpha}$, right panel), given by the model as a function of redshift, compared with observational estimates by \citet[][AS~00]{AdelbergerSteidel2000}, \citet{Ouchi2004}, \citet{Stanway2005}, \citet{Hathi2008}, \citet{Bouwens2009}, \citet{Blanc2011}, and \citet{Hayes2011}, for two bins of observed UV magnitudes: $M^{\rm obs}_{1350} \in [-22, -20]$ (blue dashed line and data points) and $M^{\rm obs}_{1350} \in [-20, -18.3]$ (red solid line and data point), roughly corresponding to LBGs and LAEs, respectively. }
}\label{fig:fesc_UVLya}
\end{figure}

\begin{figure} 
\begin{center}
	\includegraphics[width=0.45\textwidth]{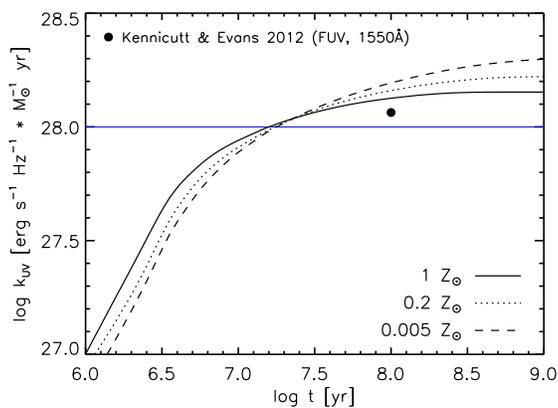}
	\caption{Evolution with galactic age of the coefficient $k_{\rm UV}$ (Equation~(\ref{eq:UV2SFR_Cha})) for a constant SFR, a \citet{Chabrier2003} IMF, and three gas metallicities: $Z_{\rm g} = 0.005$ (dashed line), 0.02 (dotted line), and 1 $Z_\odot$ (solid line). The far-UV/SFR calibration by \citet{KennicuttEvans2012} for solar metallicity and a Kroupa IMF is also shown (filled circle at $\log(t/{\rm yr})=8$). The horizontal blue line corresponds to the value adopted in the present paper.
}\label{fig:t_sfr_z}
\end{center}
\end{figure}

\begin{figure} 
\begin{center}
	\includegraphics[width=0.44\textwidth]{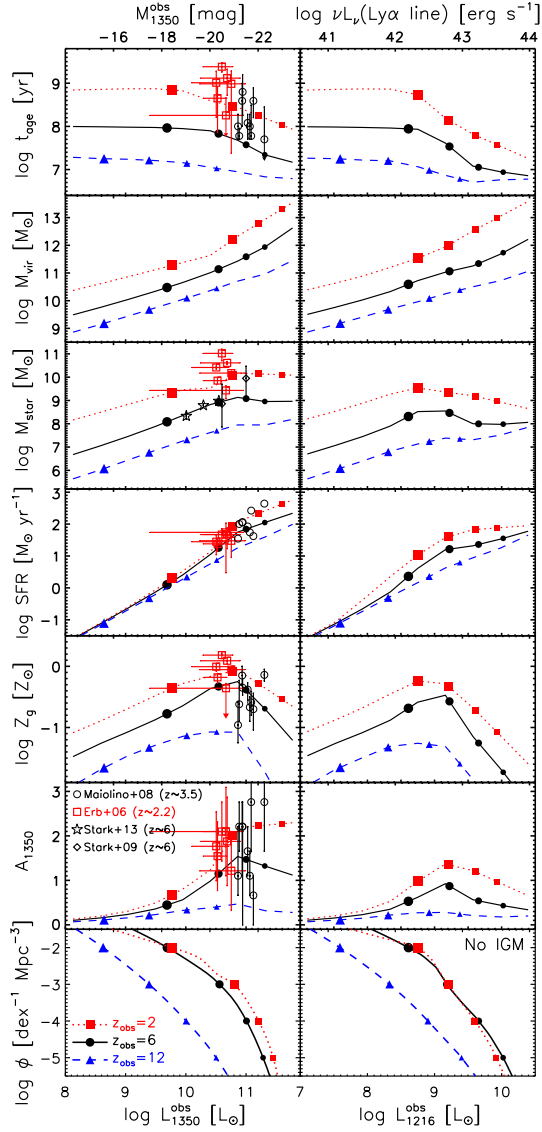}
	\caption{Average properties, weighted by the halo formation rate, of model galaxies at $z_{\rm obs} = 2$, 6, and 12  (dotted red lines and filled red squares, solid black line and black circles, dashed blue line and blue triangles, respectively) as a function of the observed UV luminosity (left panels) and of the observed Ly$\alpha$ line luminosity without the IGM attenuation (right panels). Larger symbols correspond to more numerous objects: the comoving number densities at the bin centers are $10^{-5}$, $10^{-4}$, $10^{-3}$, and $10^{-2}\,{\rm\,dex^{-1}\,Mpc^{-3}}$, respectively; in the bottom panels we can read out the corresponding luminosities. Points with error bars show observational estimates. The red open squares show the mean properties of 87 LBGs at $z \sim 2.2$ \citep{Erb2006}, the black open circles those for 9 LBGs at $z \sim 3.5$ \citep{Maiolino2008}. The stellar mass-luminosity relation at $z \sim 6$ given by \citet{Stark2009} and \citet[][broadband fluxes corrected for possible contamination of nebular emission]{Stark2013} are also shown. The legend for data symbols, given in the left $A_{1350}$ panel, applies to data in all panels.
}\label{fig:properties_LBGs_LAEs}
\end{center}
\end{figure}

\begin{figure} 
\centering
\includegraphics[width=0.6\textwidth]{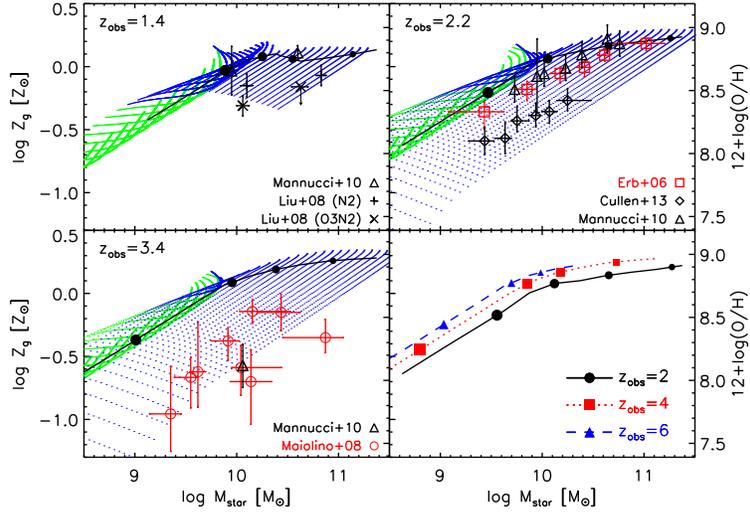}
\caption{Gas metallicity versus stellar mass at various redshifts, specified in the upper left corner of each panel. The green and blue dotted lines refer to model galaxies with halo masses of $10.5 \lesssim \log(M_{\rm vir}/M_\odot) \lesssim 11.5$ and $11.5 \lesssim \log(M_{\rm vir}/M_\odot) \lesssim 13.3$, respectively. The limited extent of the blue dotted lines at $z=1.4$ follows from the adopted lower limit to the considered virialization redshifts ($z_{\rm vir}\geqslant 1.5$), which translates into a lower limit to the stellar mass in massive halos at this redshift. The black solid lines show the average mass-metallicity relation for model galaxies, weighted by the halo formation rate. The black filled circles correspond to comoving number densities decreasing from $10^{-2}$ to $10^{-5}\,{\rm\,dex^{-1}\,Mpc^{-3}}$ in steps of one dex, with symbol size decreasing with the number density. The bottom-right panel shows the evolution of the gas metallicity versus stellar mass relation from $z_{\rm obs} = 2$ to $z_{\rm obs} = 6$. The data points are (see the legend within each panel) from \citet{Erb2006}, \citet{Liu2008}, \citet{Maiolino2008}, \citet{Mannucci2010}, and \citet{Cullen2013}.
}\label{fig:Zg_Mstar}
\end{figure}

\begin{figure*} 
\begin{center}
	\includegraphics[width=0.9\textwidth]{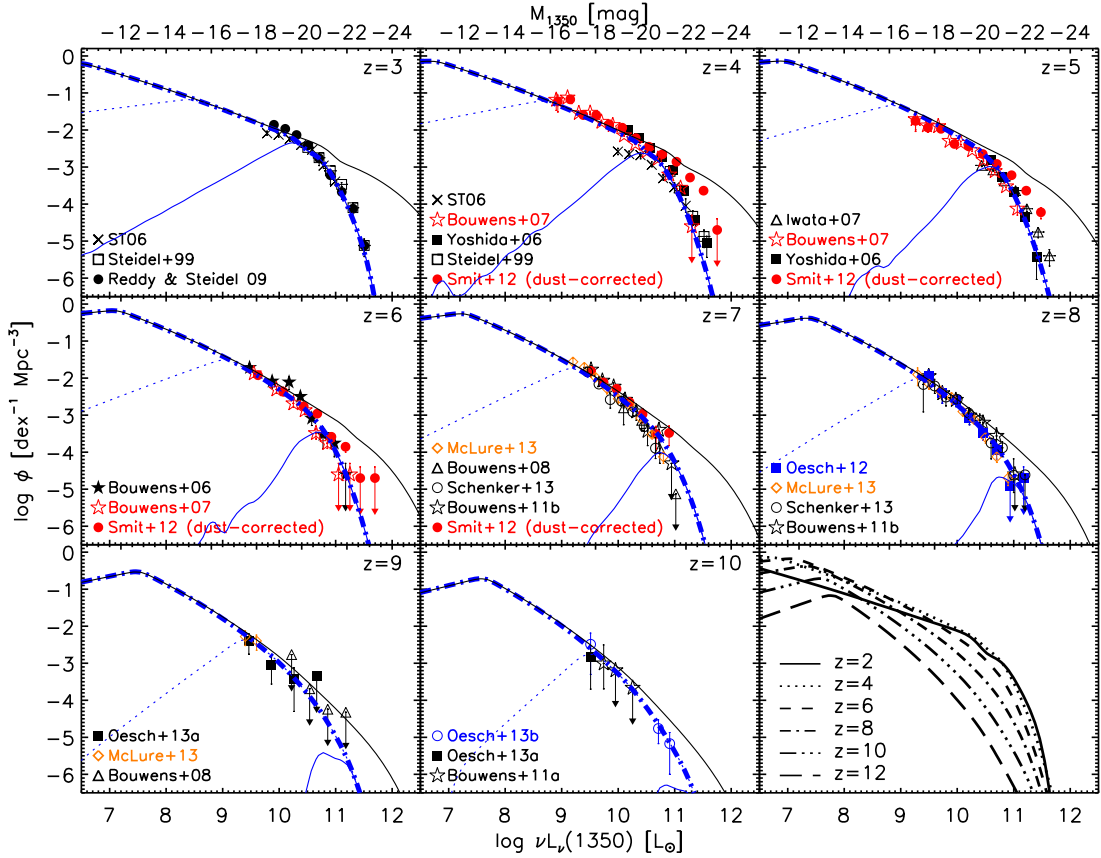}
	\caption{Luminosity functions at 1350\,$\angstrom$ at several redshifts, specified in the upper right corner of each panel.  The upper scale gives the UV magnitudes corresponding to the UV luminosities at $1350\,\angstrom$. The solid black lines show the predicted luminosity function neglecting extinction while the dot-dashed blue lines include the effect of extinction. The low luminosity break corresponds to $M_{\rm crit}=10^{8.5}\,M_\odot$. The thin dotted and solid blue lines show the effect of increasing the minimum halo mass to $10^{9.8}$ and $10^{11.2}\,M_\odot$, respectively, including extinction. The extinction affects mostly the highest luminosities, associated to the most massive objects which have the fastest chemical enrichment (see Figure~\protect\ref{fig:sph_evol_oqs1}). As illustrated by the bottom-right panel, showing the evolution of the luminosity function, its faint portion is predicted to steepen with increasing redshift. The model implies a weak evolution of the luminosity function from $z=2$ to $z=6$. The data are from \citet{Steidel1999}, \citet[ST06]{SawickiThompson2006a}, \citet{Yoshida2006}, \citet{Iwata2007}, \citet[RS09]{ReddySteidel2009}, \citet{Bouwens2006,Bouwens2007,Bouwens2008,Bouwens2011a,Bouwens2011b}, \citet{Smit2012}, \citet{Schenker2013}, \citet{McLure2013}, and \citet{Oesch2012,Oesch2013a,Oesch2013b}. Only the estimates by \citet{Smit2012} include an (uncertain) correction for dust extinction, based on the slope of the UV continuum. }\label{fig:LF_LBGs_1350}
\end{center}
\end{figure*}

\begin{figure*} 
\begin{center}
	\includegraphics[width=0.8\textwidth]{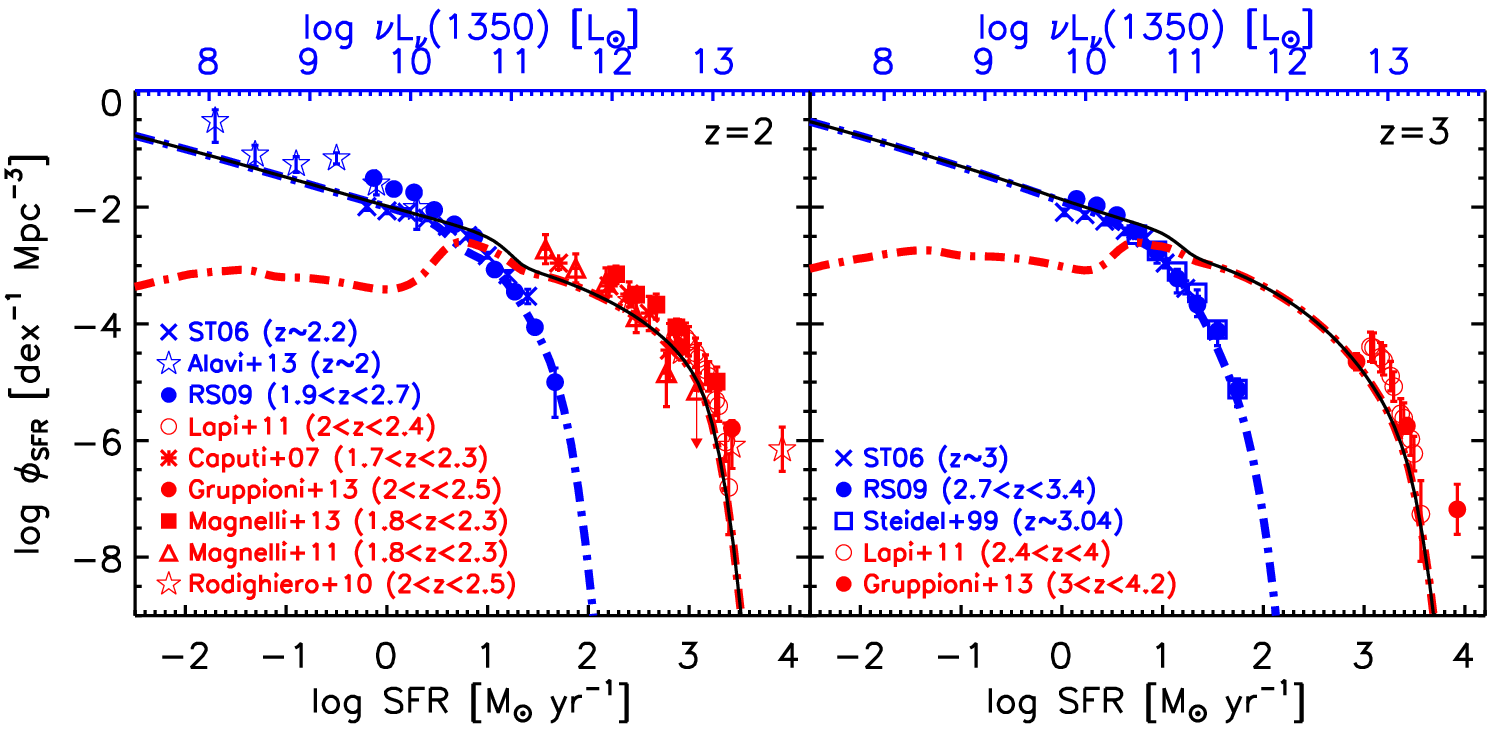}
	\caption{Comparison of the SFR functions yielded by the model at $z = 2$ and 3 with those inferred from IR \citep[8--$1000\,\mu$m, red data points;][]{Caputi2007,Rodighiero2010,Lapi2011,Gruppioni2013,Magnelli2011,Magnelli2013} and UV \citep[blue data points;][]{Steidel1999,SawickiThompson2006b,ReddySteidel2009,Alavi2014} luminosity functions. The conversion of IR luminosities into SFRs was done using the \citet{KennicuttEvans2012} calibration.  The SFR function from IR data can be directly compared to the SFR function yielded by the model (solid black line) because the star formation in these IR-bright galaxies is almost entirely dust-obscured and the contribution of older stars to dust heating is negligible. The dot-dashed blue lines show the model SFR functions as determined from Equation~(\ref{eq:UV2SFR_Cha}) with $k_{\rm UV}=1.0\times 10^{28}\,\hbox{erg}\,\hbox{s}^{-1}\,\hbox{Hz}^{-1}\,M_\odot^{-1}\,\hbox{yr}$, applied to ``observed'' (i.e., attenuated by dust) UV luminosities, and therefore these curves can be directly compared with the observed UV data, uncorrected for attenuation. The dot-dashed blue lines converge to the black lines at low SFRs, for which the dust attenuation is small.
}\label{fig:SFRF_z}
\end{center}
\end{figure*}

\begin{figure*} 
\begin{center}
	\includegraphics[width=0.9\textwidth]{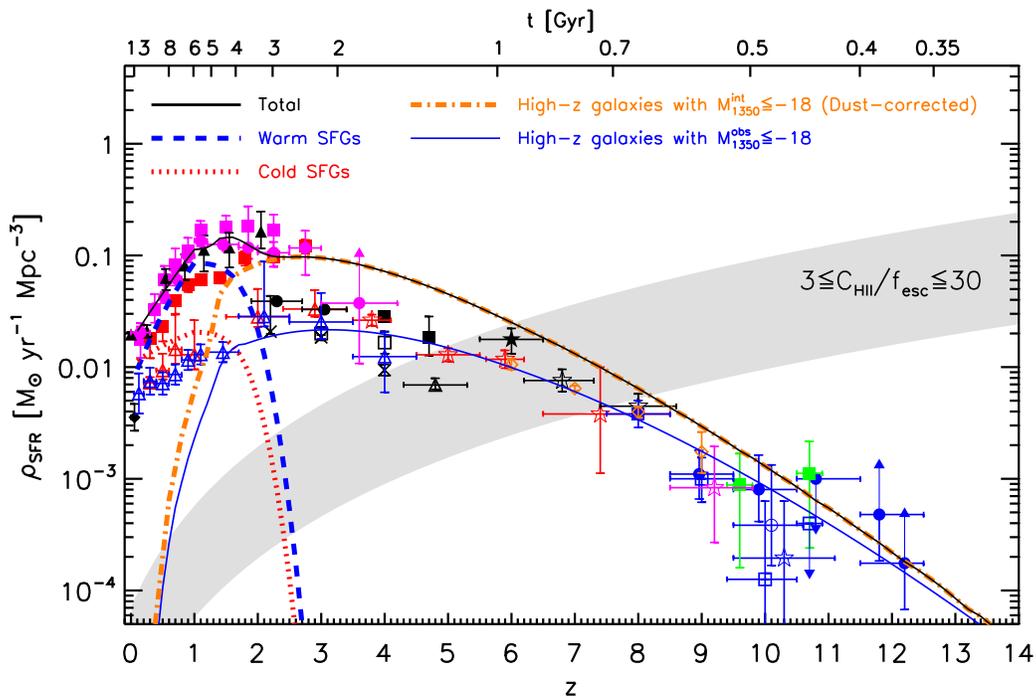}
	\caption{History of cosmic SFR density. The solid black line shows the global value, which is the sum of the contributions of warm (dashed  blue line) and cold (dotted red line) late-type galaxies and of proto-spheroidal galaxies with intrinsic $M^{\rm int}_{1350}\leqslant -18$ (dot-dashed orange line, that overlaps the solid black line for $z>2$). The SFR density of late-type galaxies was computed using the \citet{Cai2013} model; that of proto-spheroidal galaxies was computed with the present model, including  halo masses $\log(M_{\rm vir}/M_\odot) \geqslant 8.5$. The solid blue line shows the evolution, as given by our model, of the SFR density of galaxies with observed (i.e., attenuated by dust) magnitudes brighter than $M^{\rm obs}_{1350} = -18$, already represented in the observed UV luminosity functions.  The gray region illustrates the minimum SFR densities required to keep the universe fully ionized if $3 \lesssim C_{\rm HII}/f_{\rm esc} \lesssim 30$ \citep[][see also Equation~(\ref{eq:SFR_min})]{Madau1999}.
Observational estimates of SFR densities from UV data are from \citet[][filled black diamond]{Wyder2005}, \citet[][open red triangles]{Schiminovich2005}, \citet[][open blue  triangles]{Cucciati2012}, \citet[][black crosses]{SawickiThompson2006b}, \citet[][open black squares]{Steidel1999}, \citet[][filled black squares]{Yoshida2006}, \citet[][open black triangle]{Iwata2007}, \citet[][open black circles]{ReddySteidel2009}, \citet[][filled black star, open red stars, open blue star, open black stars, and open magenta star, respectively]{Bouwens2006,Bouwens2007,Bouwens2011a,Bouwens2011b,Bouwens2012a}, \citet[][filled blue square, open blue squares, open blue circle, respectively]{Oesch2012,Oesch2013a,Oesch2013b}, \citet[][filled green squares]{Coe2013}, \citet[][open orange diamonds]{McLure2013}, and \citet[][filled blue circles]{Ellis2013}. For completeness we also show SFR densities inferred from far-IR/sub-mm \citep[][filled magenta squares, filled magenta circles, and filled black triangles, respectively]{Rodighiero2010,Gruppioni2013,Magnelli2013} and radio \citep[][filled red squares]{Karim2011} data.} \label{fig:SFRD_8d5_18d0}
\end{center}
\end{figure*}


\begin{figure*} 
\begin{center}
	\includegraphics[width=0.6\textwidth]{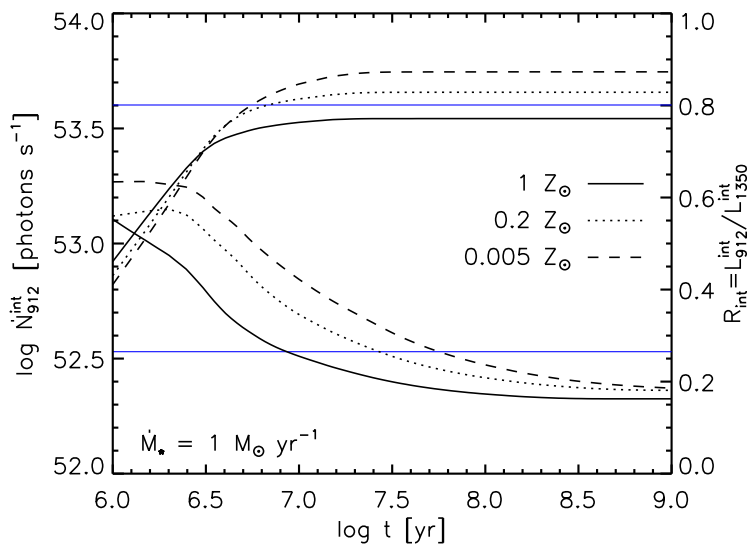}
	\caption{Evolution with galactic age of the production rate of ionizing photons, $\dot N^{\rm int}_{912}$ (left \textit{y}-scale), and of the intrinsic ratio of Lyman-continuum to UV luminosity, $R_{\rm int}\equiv L^{\rm int}_{912}/L^{\rm int}_{1350}$ (right \textit{y}-scale) for a constant SFR, $\dot M_\star = 1\ M_\odot\ \rm yr^{-1}$, a \citet{Chabrier2003} IMF, and three metallicities: $Z_{\rm g} = 0.005$ (dashed line), 0.02 (dotted line), and 1 $Z_\odot$ (solid line). The chosen reference values, $\dot N^{\rm int}_{912} = 4.0 \times 10^{53}$ and $R_{\rm int}=0.265$, are indicated by the upper and lower horizontal lines, respectively.
}\label{fig:kion_Rint}
\end{center}
\end{figure*}


\begin{figure*} 
\begin{center}
	\includegraphics[width=0.9\textwidth]{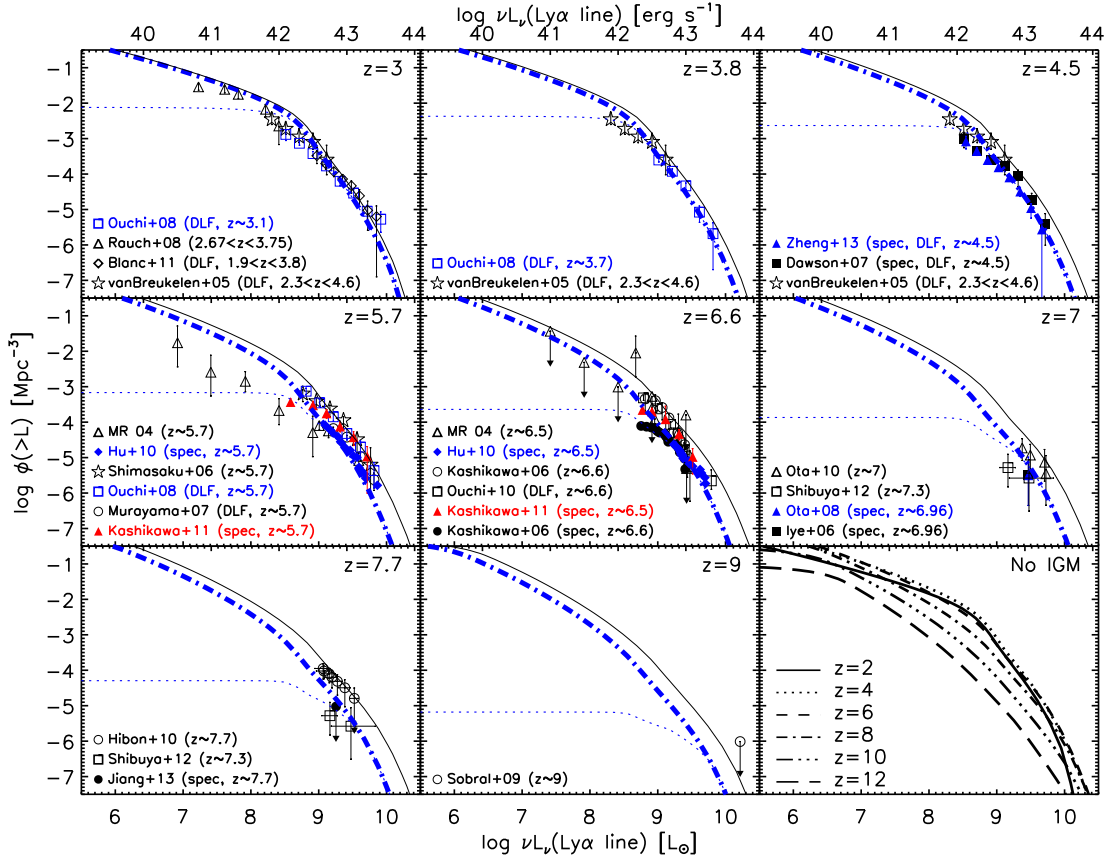}
	\caption{Model cumulative Ly$\alpha$ line luminosity functions at several redshifts, specified in the upper right corner of each panel corrected (solid black lines) and uncorrected (dot-dashed blue lines) for attenuation by the IGM computed following \citet{Madau1995}. We have  adopted a minimum halo mass of $10^{8.5}\,M_\odot$. The dotted blue lines give the contributions of halo masses $\geqslant 10^{11}\,M_\odot$. The bottom-right panel illustrates the evolution of the Ly$\alpha$ line luminosity function, without attenuation by the IGM. The data are from \citet[]{Ouchi2003}, \citet[][MR 04]{MalhotraRhoads2004}, \citet[]{vanBreukelen2005}, \citet[]{Shimasaku2006}, \citet[]{Iye2006}, \citet[]{Kashikawa2006,Kashikawa2011}, \citet[]{Dawson2007}, \citet[]{Murayama2007}, \citet[]{Rauch2008}, \citet[]{Ouchi2008,Ouchi2010}, \citet[]{Ota2008,Ota2010}, \citet[]{Sobral2009}, \citet[]{Hibon2010}, \citet[]{Hu2010}, \citet[]{Blanc2011}, \citet[]{Shibuya2012}, \citet[]{Zheng2013}, and \citet[]{Jiang2013}. The data based on spectroscopic samples are shown by filled symbols and the references within panels are labeled ``spec''. Those based on photometric samples are shown by open symbols. The label ``DLF'' associated to references means that the original papers gave the differential luminosity functions. }\label{fig:LF_LAEs_c}
\end{center}
\end{figure*}


\begin{figure*} 
\begin{center}
	\includegraphics[width=0.7\textwidth]{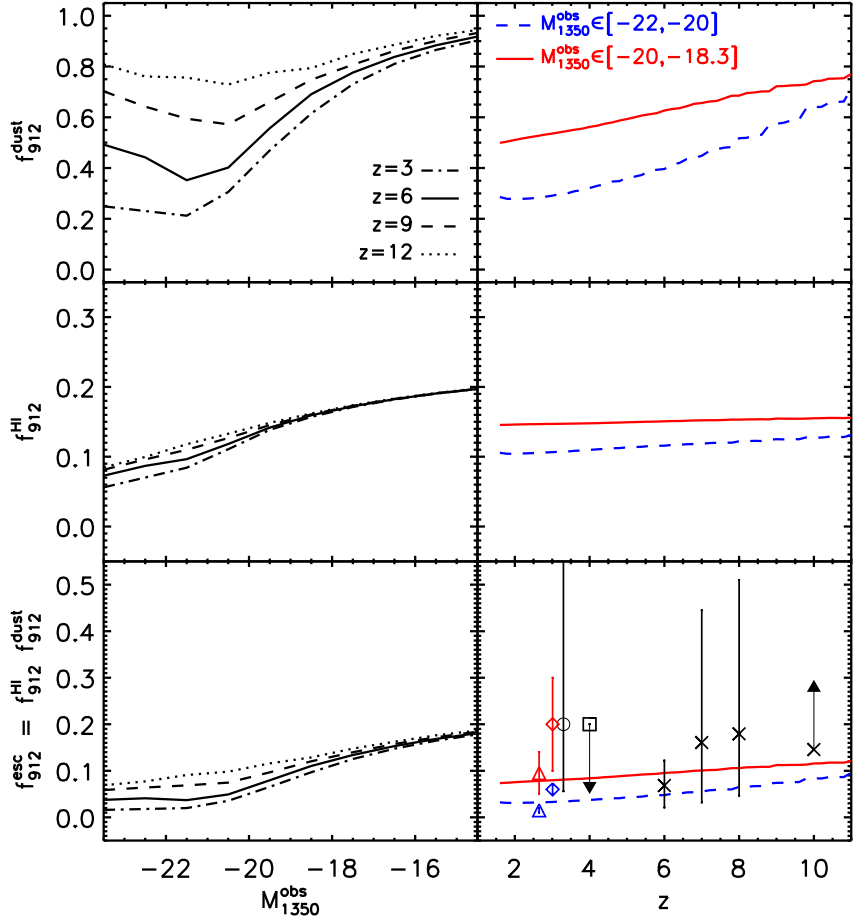}
	\caption{{\textit{Left panels}: fractions of ionizing photons surviving dust, HI, and both absorptions ($f^{\rm dust}_{912}$, $f^{\rm HI}_{912}$, and $f^{\rm esc}_{912}$, from top to bottom), weighted by the halo formation rate, yielded by our model for $z_{\rm obs} = 3$ (dot-dashed line), 6 (solid line), 9 (dashed line), and 12 (dotted line) as a function of the attenuated UV luminosity $M^{\rm obs}_{1350}$. \textit{Right panels}: escape fractions given by the model as a function of $z$ for two luminosity bins, i.e., $M^{\rm obs}_{1350} \in [-22, -20]$ (dashed blue line) and $M^{\rm obs}_{1350} \in [-20, -18.3]$ (solid red line). Data in the bottom-right panel are from: \citet{Iwata2009} at $z \simeq 3.1$ (offset by $\Delta z = 0.2$ for readability; the open circle corresponds to the median value and the error bars extend to the minimum/maximum values); \citet{Nestor2013} for LBGs ($M_{\rm UV} \in [-22, -20]$; open blue diamond) and for LAEs ($M_{\rm UV} \in [-20, -18.3]$; open red diamond) at $z \sim 3$; \citet[][open blue triangle and open red triangle for LBGs and LAEs, respectively]{Mostardi2013}  at $z \sim 2.85$ (the points are offset by $\Delta z = -0.2$ for readability); \citet[][open square representing an upper limit at $z \simeq 4$]{Vanzella2010}; \citet[][black crosses]{Mitra2013} for LBGs in the redshift range $6 \leqslant z \leqslant 10$. }
}\label{fig:fesc_Mz2}
\end{center}
\end{figure*}

\begin{figure*} 
\begin{center}
	\includegraphics[width=0.6\textwidth]{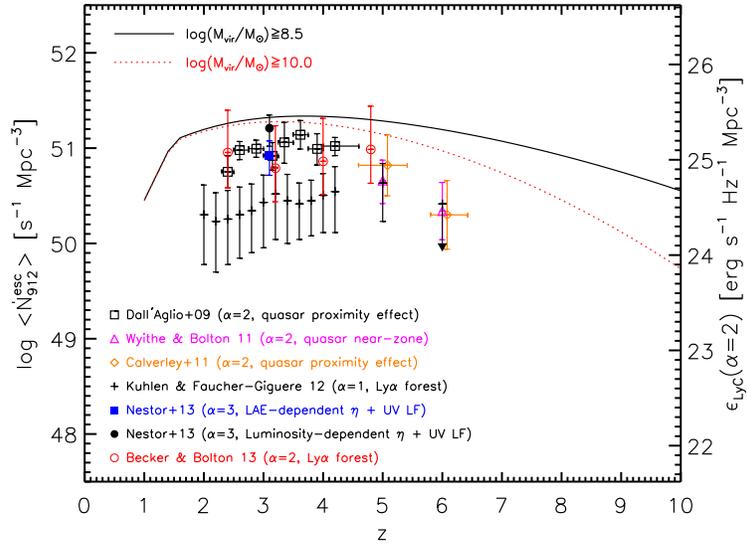}
	\caption{Comoving emission rates of ionizing photons ($\langle \dot N^{\rm esc}_{912} \rangle$, left $y$-scale) and ionizing emissivities ($\epsilon_{\rm LyC} \simeq \langle \dot N^{\rm esc}_{912} \rangle h_{\rm P} \alpha$ for $\epsilon(\nu) = \epsilon_{\rm LyC} (\nu/\nu_{912})^{-\alpha}$ with $\alpha = 2$, right $y$-scale) as a function of redshift. The solid black line and the dotted red line correspond to critical halo masses of $10^{8.5}$ and $10^{10}\,M_\odot$, respectively. Data points are from \citet{DallAglio2009}, \citet{WyitheBolton2011}, \citet{Calverley2011}, \citet{KuhlenFaucherGiguere2012}, \citet{BeckerBolton2013}, and \citet{Nestor2013}.
}\label{fig:dN912dtdV}
\end{center}
\end{figure*}

\begin{figure*} 
\begin{center}
	\includegraphics[width=\textwidth]{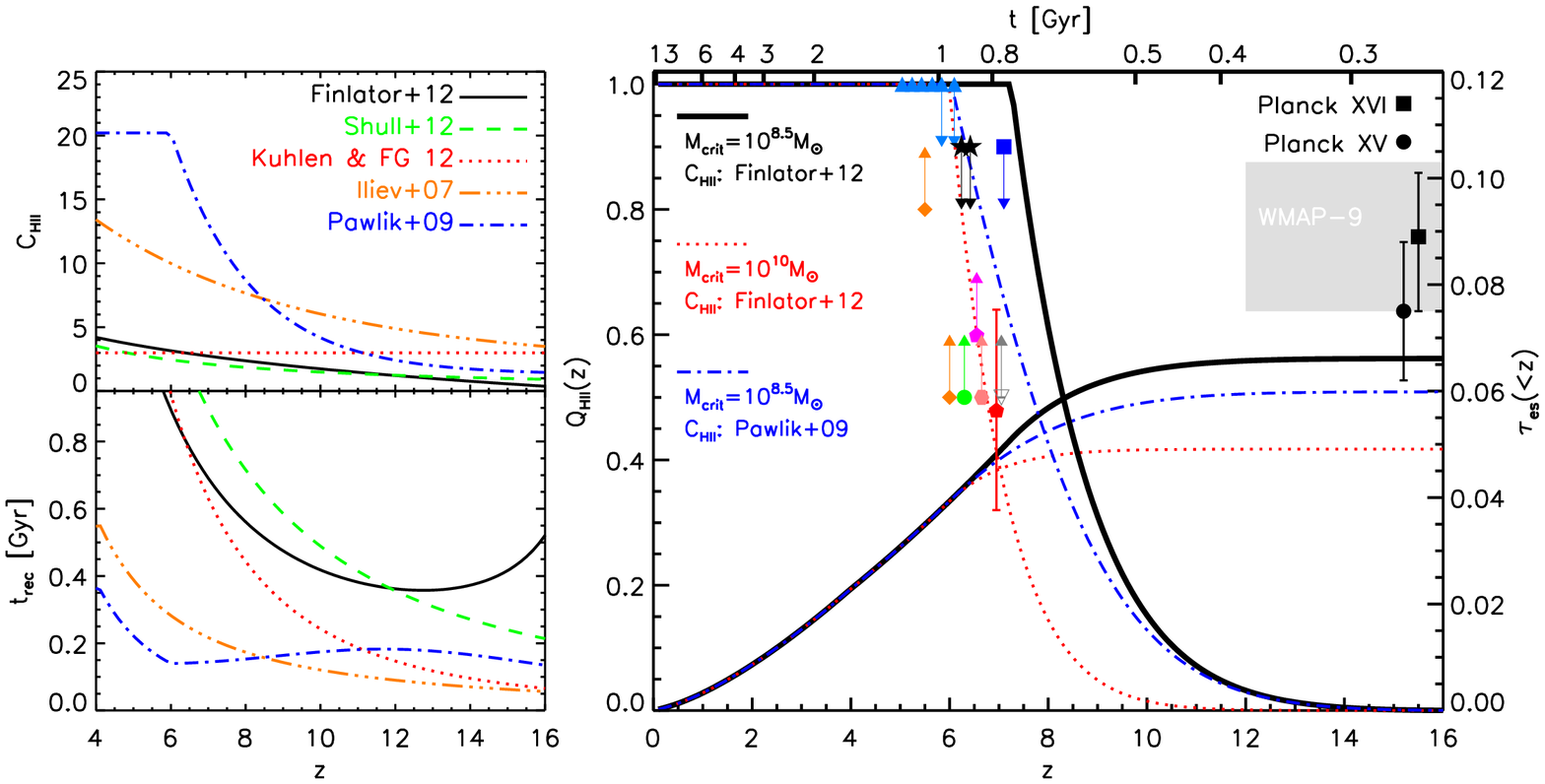}
\caption{\textit{Left panels}: evolutionary laws for the IGM clumping factor $C_{\rm HII}$ (upper panel) proposed in the literature and the corresponding recombination timescales $\bar t_{\rm rec}$ (lower panel). Solid black line: $C_{\rm HII}(z) = 9.25 - 7.21 \log(1 + z)$ \citep{Finlator2012}; dashed green line: $C_{\rm HII}(z) = 2.9 [(1+z)/6]^{-1.1}$ \citep{Shull2012}; dotted red line: $C_{\rm HII}(z) = 3$ \citep{KuhlenFaucherGiguere2012}; triple-dot-dashed orange line: $C_{\rm HII}(z) = 26.2917 \exp(-0.1822z+0.003505z^2)$ \citep{Iliev2007}; dot-dashed blue line: $C_{\rm HII}(z) = \min[C_{\rm HII}(z=6), \exp(-0.47z+5.76)+1.29]$, corresponding to the $C_{-1}$ L6N256 no-reheating simulation by \citet{Pawlik2009} covering the range $6 \leqslant z \leqslant 20$.
\textit{Main figure}: evolution with redshift of the volume filling factor $Q_{\rm HII}$ (left \textit{y}-scale) and of the electron optical depth $\tau_{\rm es}$ (right \textit{y}-scale). The thick solid black lines correspond to the fiducial model with $C_{\rm HII}(z)$ by \citet{Finlator2012} and $M_{\rm crit}=10^{8.5}\,M_\odot$. The dotted red lines correspond to the same model but with $M_{\rm crit} = 10^{10}\ M_\odot$. The dot-dashed blue lines show the results with the $C_{\rm HII}(z)$ by \citet{Pawlik2009} for $M_{\rm crit}=10^{8.5}\,M_\odot$. The observational constraints on the volume filling factor are from a collection of literature data made by \citet{Robertson2013}. The 9-year \textit{WMAP} constraint on electron optical depth, $\tau_{\rm es} = 0.089 \pm 0.014$  \citep{Hinshaw2013}, is represented by the gray region while the filled square and the filled circle with error bars represent the preliminary estimates by \citet{PlanckCollaborationXVI2013} and \citet{PlanckCollaborationXV2013}, respectively.
}\label{fig:Reion_tau}
\end{center}
\end{figure*}



\end{document}